\title{Measuring the gravitational field in General Relativity: From deviation equations and the gravitational compass to relativistic clock gradiometry}

\author{Yuri N.\ Obukhov\footnote{Email: obukhov@ibrae.ac.ru}\\
        Theoretical Physics Laboratory\\
        Nuclear Safety Institute\\
        Russian Academy of Sciences, B.Tulskaya 52, 115191 Moscow, Russia \and \\
        Dirk Puetzfeld\footnote{Email: dirk.puetzfeld@zarm.uni-bremen.de, URL: http://puetzfeld.org} \\
        University of Bremen\\
				Center of Applied Space Technology and Microgravity (ZARM)\\
				28359 Bremen, Germany
        }
\date{\today}

\documentclass[11pt,english]{article}
\usepackage[]{graphicx}
\usepackage{amsmath,amsfonts,amssymb}

\begin{document}

\def\ut{{\underset {\widetilde{\ \ }}u}}
\def\at{{\underset {\widetilde{\ \ }}a}}
\def\ap{\check{a}}

\maketitle

\begin{abstract}
How does one measure the gravitational field? We give explicit answers to this fundamental question and show how all components of the curvature tensor, which represents the gravitational field in Einstein's theory of General Relativity, can be obtained by means of two different methods. The first method relies on the measuring the accelerations of a suitably prepared set of test bodies relative to the observer. The second methods utilizes a set of suitably prepared clocks. The methods discussed here form the basis of relativistic (clock) gradiometry and are of direct operational relevance for applications in geodesy.
\end{abstract}
\newpage

\tableofcontents

\section{Introduction}\label{sec_introduction}

The measurement of the gravitational field lies at the heart of gravitational physics and geodesy. Here we provide the relativistic foundation and present two methods for the operational determination of the gravitational field.

In Einstein's theory of gravitation, i.e.\ General Relativity (GR)\index{General Relativity}, the gravitational field manifests itself in the form of the Riemannian {curvature tensor}\index{tensor!curvature} $R_{abc}{}^{d}$ \cite{Pirani:1956,Synge:1960}. This 4th-rank tensor can be defined as a measure of the noncommutativity of the parallel transport process of the underlying spacetime manifold $M$. In terms of the {covariant derivative}\index{derivative!covariant} $\nabla_a$, and for a mixed tensor $T^{c}{}_{d}$, it is introduced via
\begin{eqnarray}
\left(\nabla_a\nabla_b - \nabla_b\nabla_a\right)T^{c}{}_{d} 
= R_{abe}{}^{c}\,T^{e}{}_{d} - R_{abd}{}^{e}\,T^{c}{}_{e}. \label{curvature_def}
\end{eqnarray}
Note that our curvature conventions differ from those in \cite{Synge:1960,Poisson:etal:2011}. General Relativity is formulated on a four-dimensional (pseudo) Riemannian spacetime\index{spacetime!Riemannian} with the metric $g_{ab}$ of the signature $(+1,-1,-1,-1)$ which is compatible with the connection in the sense of $\nabla_cg_{ab} = 0$. Therefore the curvature tensor in Einstein's theory has twenty (20) independent components for the most general field configurations produced by nontrivial matter sources, whereas in vacuum the number of independent components reduces to ten (10). As compared to Newton's theory, the  gravitational field thus has more degrees of freedom in the relativistic framework. 

The smooth tensor field $g_{ab}(x^c)$ introduces the metricity relations on the spacetime manifold $M$: an interval (``distance'') between any two close points $x\in M$ and $x + dx\in M$ is defined by
\begin{equation}
ds^2 = g_{ab}\,dx^adx^b.\label{ds2}
\end{equation}
The metric and connection $(g, \nabla)$ underlie the formalism of Synge's world function \cite{Synge:1960} which plays a crucial role in the methods of measurement of the gravitational field in GR and in its natural extensions. 

A central question in General Relativity, and consequently in relativistic geodesy, is how these components of the {gravitational field}\index{gravitational! field} can be determined in an operational way. 

\subsection{Method 1: Measuring the gravitational field by means of test bodies}\label{subsec_method_1} 

Method 1 utilizes a suitably prepared set of test bodies in order to determine all components of the curvature of spacetime and thereby the gravitational field. This method relies on the measurement of the acceleration between the test-bodies and the observer. Historically, Felix Pirani \cite{Pirani:1956} was the first to point out that one could determine the full Riemann tensor\index{tensor!Riemann curvature} with the help of a (sufficiently large) number of test bodies in the vicinity of observer's world line. Pirani's suggestion to measure the curvature was based on the equation which describes the dynamics of a vector connecting two adjacent geodesics in spacetime. In the literature this equation is known as a {Jacobi equation}\index{equation!Jacobi}, or a {geodesic deviation equation}\index{equation!geodesic deviation}; its early derivations in a Riemannian context can be found in \cite{LeviCivita:1926,Synge:1926,Synge:1927}. 

\begin{figure}
  \begin{minipage}[c]{0.6\textwidth}
    \includegraphics[width=\textwidth,angle=-90]{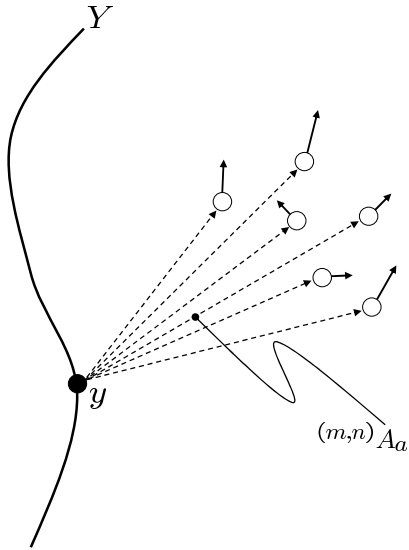}
  \end{minipage}\hfill
  \begin{minipage}[c]{0.4\textwidth}
    \caption{\label{fig_compass_sketch} Method 1: Sketch of the operational procedure to measure the curvature of spacetime. An observer moving along a world line $Y$ monitors the accelerations ${}^{(m,n)}A_a$ to a set of suitably prepared test bodies (hollow circles). The number of test bodies required for the determination of all curvature components depends on the type of the underlying spacetime.} 
  \end{minipage}
\end{figure}

A modern derivation and extension of the deviation equation, based on \cite{Puetzfeld:Obukhov:2016:1}, is presented in the next section. In particular, it is explicitly shown, how a suitably prepared set of test bodies can be used to determine all components of the curvature of spacetime (and thereby to measure the gravitational field) with the help of an exact solution for the components of the Riemann tensor in terms of the mutual accelerations between the constituents of a cloud of test bodies and the observer.\index{observer} This can be viewed as an explicit realization of Szekeres' ``{gravitational compass}\index{gravitational!compass}'' \cite{Szekeres:1965}, or Synge's ``curvature detector'' \cite{Synge:1960}. In geodetic terms, such a solution represents a realization of a {relativistic gradiometer}\index{gradiometer!relativistic} or {tensor gradiometer}\index{gradiometer!tensor}, which has a direct operational relevance and forms the basis of relativistic gradiometry. The operational procedure, see fig.\ \ref{fig_compass_sketch}, is to monitor the accelerations of a set of test bodies w.r.t.\ to an observer moving along a reference world line $Y$. A mechanical analogue would be to measure the forces between the test bodies and the reference body via a spring connecting them. 

Method 1 relies on the standard geodesic deviation equation. A modern covariant derivation of this equation, as well as its generalization to higher orders will be provided to make the presentation self-contained. 
Furthermore, we provide an explicit exact solution for the curvature components in terms of the mutual accelerations between the constituents of a cloud of test bodies and the observer. Our presentation is mainly based on \cite{Puetzfeld:Obukhov:2016:1}.   

\subsection{Method 2: Measuring the gravitational field by means of clocks}\label{subsec_method_2}

Method 2 utilizes a suitably prepared set of clocks to determine all components of the gravitational field in General Relativity. In contrast to the gravitational compass, the method relies on the frequency comparison between the clocks from the ensemble and the one carried by the observer. We call such an experimental setup a clock compass, in analogy to the usual gravitational compass, or in geodetic language a clock gradiometer.  

We base our review on \cite{Puetzfeld:Obukhov:2018:1} and pay particular attention to the construction of the underlying reference frame. As in the case of the gravitational compass, our results are of direct operational relevance for the setup of networks of clocks, for example in the context of relativistic geodesy.

\begin{figure}
  \begin{minipage}[c]{0.6\textwidth}
    \includegraphics[width=\textwidth,angle=-90]{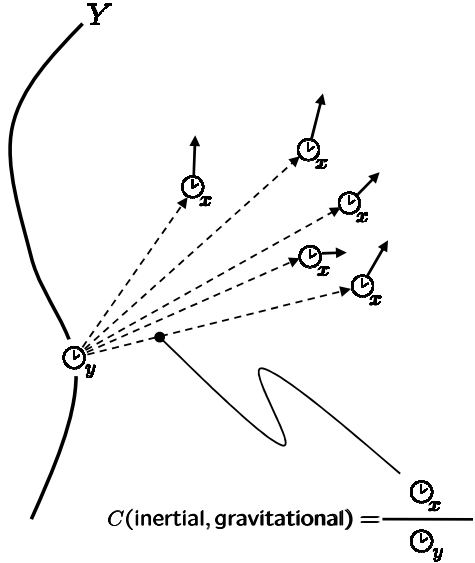}
  \end{minipage}\hfill
  \begin{minipage}[c]{0.4\textwidth}
    \caption{\label{fig_clock_compass_sketch} Method 2: Sketch of the operational procedure to measure the curvature of spacetime. An observer with a clock moving along a world line $Y$ compares his clock readings $C$ to a set of suitably prepared clocks in the vicinity of $Y$. The number of clocks required for the determination of all curvature components depends on the type of the underlying spacetime.} 
  \end{minipage}
\end{figure}

\section{Theoretical foundations}\label{sec_theoretical_foundations}

In this section we present the theoretical foundations for method 1 and method 2. 

For method 1 we start by comparing two general curves in an arbitrary spacetime manifold and work out an equation for the generalized deviation vector between those two curves in section \ref{subsec_theo_found_1}.

For method 2 we first show in section \ref{subsec_theo_found_2} how the metric along an arbitrary world line can be expressed in terms of geometrical and kinematical parameters. This result is then used in section \ref{subsec_theo_found_3} to derive the frequency ratio of two clocks moving on two general curves, again within an arbitrary spacetime manifold.

\subsection{Method 1: Comparison of two general curves}\label{subsec_theo_found_1}

Consider two curves $Y(t)$ and $X(\tilde{t})$ with general parameters $t$ and $\tilde{t}$, i.e.\ are not necessarily the proper time on the given curves. Now we connect two points $x\in X$ and $y\in Y$ on the two curves by the geodesic joining the two points (we assume that this geodesic is unique). 

For the geodesic connecting the two general curves $Y(t)$ and $X(\tilde{t})$ we have the world function introduced as an integral
\begin{equation}
\sigma(x,y) := \frac{\epsilon}{2}\Biggl( \int\limits_x^y d\tau \Biggr)^2\label{world}
\end{equation}
over the geodesic curve $\gamma$ connecting the spacetime points $x$ and $y$. Here $d\tau = \sqrt{g_{ab}u^au^b}\,d\lambda$ is the differential of the proper time along the geodesic which is defined as the curve $\gamma = \{x^a(\lambda)\}$, such that the tangent vector $u^a = dx^a/d\lambda$ is parallely transported $Du^a/d\lambda = 0$, and $\epsilon = \pm 1$ for timelike/spacelike curves. Along with the world function $\sigma(x,y)$, another important bitensor is the parallel propagator $g^y{}_x(x,y)$ that allows for the parallel transportation of objects along the unique geodesic that links the points $x$ and $y$. For example, given a vector $V^x$ at $x$, the corresponding vector at $y$ is obtained by means of the parallel transport along the geodesic curve as $V^y = g^y{}_x(x,y)V^x$. For more details see, e.g., \cite{Synge:1960,DeWitt:Brehme:1960} or section 5 in \cite{Poisson:etal:2011}. A compact summary of useful formulas in the context of the bitensor formalism can also be found in the appendices A and B of \cite{Puetzfeld:Obukhov:2013}. Note that we will use the condensed notation when the spacetime point to which an index of a bitensor belongs can be directly read from an index itself. Indices attached to the world-function always denote covariant derivatives, at the given point, i.e.\ $\sigma_y:= \nabla_y \sigma$, hence we do not make explicit use of the semicolon in case of the world-function.

\begin{figure}
  \begin{minipage}[c]{0.55\textwidth}
    \includegraphics[width=\textwidth]{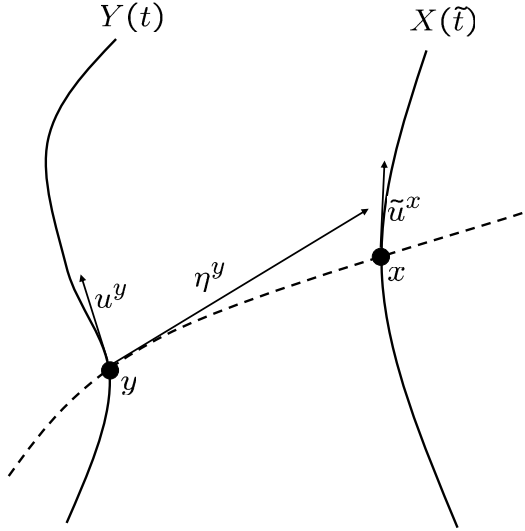}
  \end{minipage}\hfill
  \begin{minipage}[c]{0.4\textwidth}
    \caption{\label{fig_setup_method_1} Sketch of the two arbitrarily parametrized world lines $Y(t)$ and $X(\tilde{t})$, and the geodesic connecting two points on these world line. The (generalized) deviation vector along the reference world line $Y$ is denoted by $\eta^y$.}
  \end{minipage}
\end{figure}

Conceptually, the closest object to the connecting vector between the two points is the covariant derivative of the world function: $\sigma^y$. Note though that $\sigma^y$ is just tangent at that point (its length being the the geodesic length between $y$ and $x$), only in flat spacetime it coincides with the connecting vector. Keeping in mind such an interpretation, let us now work out a propagation equation for this ``generalized'' connecting vector along the reference curve, cf.\ fig.\ \ref{fig_setup_method_1}. Following our conventions the reference curve will be $Y(t)$ and we define the generalized connecting vector to be:
\begin{eqnarray}
\eta^y := - \sigma^y\,. \label{gen_dev_definition} 
\end{eqnarray}
Taking its covariant total derivative, we have
\begin{eqnarray}
\frac{D}{dt} \eta^{y_1} &=& - \frac{D}{dt} \sigma^{y_1}\left(Y(t),X(\tilde{t})\right) \nonumber \\
&=& - \sigma^{y_1}{}_{y_2} \frac{\partial Y^{y_2}}{\partial t} - \sigma^{y_1}{}_{x_2} \frac{\partial X^{x_2}}{\partial \tilde{t}} \frac{d\tilde{t}}{dt} \nonumber \\
&=& - \sigma^{y_1}{}_{y_2} u^{y_2} - \sigma^{y_1}{}_{x_2} \tilde{u}^{x_2} \frac{d\tilde{t}}{dt}, \label{eta_1st_deriv}
\end{eqnarray}
where in the last line we defined the velocities along the two curves $Y$ and $X$. As usual, $\sigma^y{}_{x_1\dots y_2\dots} := \nabla_{x_1}\dots\nabla_{y_2}\dots (\sigma^y)$ denote the higher order covariant derivatives of the world function. We continue by taking the second derivative of (\ref{eta_1st_deriv}), which yields
\begin{eqnarray}
\frac{D^2}{dt^2} \eta^{y_1} &=& - \sigma^{y_1}{}_{y_2 y_3} u^{y_2} u^{y_3} - 2 \sigma^{y_1}{}_{y_2 x_3} u^{y_2} \tilde{u}^{x_3} \frac{d\tilde{t}}{dt} \nonumber \\
&& - \sigma^{y_1}{}_{y_2} a^{y_2} - \sigma^{y_1}{}_{x_2 x_3} \tilde{u}^{x_2} \tilde{u}^{x_3} \left(\frac{d\tilde{t}}{dt} \right)^2 \nonumber \\
&& - \sigma^{y_1}{}_{x_2} \tilde{a}^{x_2} \left(\frac{d\tilde{t}}{dt} \right)^2 - \sigma^{y_1}{}_{x_2} \tilde{u}^{x_2}  \frac{d^2\tilde{t}}{dt^2}, \label{eta_2nd_deriv}
\end{eqnarray}
here we introduced the accelerations $a^y:={D u^y}/dt$, and $\tilde{a}^x:={D \tilde{u}^x}/d\tilde{t}$. Equation (\ref{eta_2nd_deriv}) is already the generalized deviation equation, but the goal is to have all the quantities therein defined along the reference wordline $Y$. 

%
%\begin{figure}
%\begin{center}
%\includegraphics[width=7cm]{deviation_fig1.eps}
%\end{center}
%\caption{\label{fig_setup_method_1} Sketch of the two arbitrarily parametrized world lines $Y(t)$ and $X(\tilde{t})$, and the geodesic connecting two points on these world line. The (generalized) deviation vector along the reference world line $Y$ is denoted by $\eta^y$.}
%\end{figure}

We now derive some auxiliary formulas, by introducing the inverse of the second derivative of the world function via the following equations:
\begin{eqnarray}
\stackrel{-1}{\sigma}{}\!\!^{y_1}{}_x \sigma^{x}{}_{y_2} = \delta^{y_1}{}_{y_2},\qquad 
\stackrel{-1}{\sigma}{}\!\!^{x_1}{}_y \sigma^{y}{}_{x_2} = \delta^{x_1}{}_{x_2}. \label{inverse_2}
\end{eqnarray}
Multiplication of (\ref{eta_1st_deriv}) by $\stackrel{-1}{\sigma}{}\!\!^{x_3}{}_{y_1}$ then yields
\begin{eqnarray}
\tilde{u}^{x_3} \frac{d\tilde{t}}{dt} &=& - \stackrel{-1}{\sigma}{}\!\!^{x_3}{}_{y_1} \sigma^{y_1}{}_{y_2} u^{y_2} + \stackrel{-1}{\sigma}{}\!\!^{x_3}{}_{y_1}  \frac{D \sigma^{y_1}}{dt} \nonumber \\ 
&=& K^{x_3}{}_{y_2} u^{y_2} - H^{x_3}{}_{y_1} \frac{D \sigma^{y_1}}{dt}. \label{formal_vel_1}
\end{eqnarray}
In the last line we defined two auxiliary quantities $K^x{}_{y}$ and $H^x{}_{y}$ -- the notation follows the terminology of Dixon. Equation (\ref{formal_vel_1}) allows us to formally express the the velocity along the curve $X$ in terms of the  quantities which are defined at $Y$ and then ``propagated'' by $K^x{}_{y}$ and $H^x{}_{y}$. Using (\ref{formal_vel_1}) in (\ref{eta_2nd_deriv}) we arrive at:
\begin{eqnarray}
\frac{D^2}{dt^2} \eta^{y_1} &=& - \sigma^{y_1}{}_{y_2 y_3} u^{y_2} u^{y_3} - \sigma^{y_1}{}_{y_2} a^{y_2} \nonumber \\
&&- 2 \sigma^{y_1}{}_{y_2 x_3} u^{y_2} \left( K^{x_3}{}_{y_4} u^{y_4} - H^{x_3}{}_{y_4} \frac{D \sigma^{y_4}}{dt} \right) \nonumber \\
&& - \sigma^{y_1}{}_{x_2 x_3}  \left( K^{x_2}{}_{y_4} u^{y_4} - H^{x_2}{}_{y_4} \frac{D \sigma^{y_4}}{dt} \right) \nonumber \\
&& \times \left( K^{x_3}{}_{y_5} u^{y_5} - H^{x_3}{}_{y_5} \frac{D \sigma^{y_5}}{dt} \right) \nonumber \\
&& - \sigma^{y_1}{}_{x_2} \frac{D}{dt} \left( K^{x_2}{}_{y_3} u^{y_3} - H^{x_2}{}_{y_3} \frac{D \sigma^{y_3}}{dt} \right). \label{eta_2nd_deriv_alternative_1}
\end{eqnarray}
We may derive an alternative version of this equation -- by using (\ref{formal_vel_1}) multiplied by $d t / d \tilde{t}$ -- which yields
\begin{eqnarray}
\tilde{u}^{x_3} &=& K^{x_3}{}_{y_2} u^{y_2} \frac{dt}{d\tilde{t}} - H^{x_3}{}_{y_1} \frac{D \sigma^{y_1}}{dt} \frac{dt}{d\tilde{t}}, \label{formal_vel_2}
\end{eqnarray} 
and inserted into (\ref{eta_2nd_deriv}):
\begin{eqnarray}
\frac{D^2}{dt^2} \eta^{y_1} &=& - \sigma^{y_1}{}_{y_2 y_3} u^{y_2} u^{y_3} - \sigma^{y_1}{}_{y_2} a^{y_2} - \sigma^{y_1}{}_{x_2} \tilde{a}^{x_2} \left(\frac{d\tilde{t}}{dt} \right)^2  
\end{eqnarray}
\begin{eqnarray}
&&- 2 \sigma^{y_1}{}_{y_2 x_3} u^{y_2} \left( K^{x_3}{}_{y_4} u^{y_4} - H^{x_3}{}_{y_4} \frac{D \sigma^{y_4}}{dt} \right) \nonumber \\
&& - \sigma^{y_1}{}_{x_2 x_3}  \left( K^{x_2}{}_{y_4} u^{y_4} - H^{x_2}{}_{y_4} \frac{D \sigma^{y_4}}{dt} \right) \nonumber \\
&& \times \left( K^{x_3}{}_{y_5} u^{y_5} - H^{x_3}{}_{y_5} \frac{D \sigma^{y_5}}{dt} \right) \nonumber \\
&& - \sigma^{y_1}{}_{x_2} \frac{d t}{d \tilde{t}} \frac{d^2\tilde{t}}{dt^2} \left( K^{x_2}{}_{y_3} u^{y_3} - H^{x_2}{}_{y_3} \frac{D \sigma^{y_3}}{dt} \right). \label{eta_2nd_deriv_alternative_2}
\end{eqnarray}
Note that we may determine the factor $d \tilde{t} / dt $ by requiring that the velocity along the curve $X$ is normalized, i.e.\ $\tilde{u}^x \tilde{u}_x =1$, in which case (\ref{formal_vel_1}) yields
\begin{eqnarray}
\frac{d\tilde{t}}{dt} &=& \tilde{u}_{x_1} K^{x_1}{}_{y_2} u^{y_2} -  \tilde{u}_{x_1} H^{x_1}{}_{y_2} \frac{D \sigma^{y_2}}{dt}. \label{formal_vel_3}
\end{eqnarray} 

\paragraph{Expansion of quantities on the reference world line}

The generalized (exact) deviation equations (\ref{eta_2nd_deriv_alternative_1}) and (\ref{eta_2nd_deriv_alternative_2}) contain quantities which are not defined along the reference curve, in particular the covariant derivatives of the world function. We now make use of the covariant expansions of these quantities, which read (for details, see \cite{Puetzfeld:Obukhov:2014:2}): 
\begin{eqnarray}
\sigma^{y_0}{}_{x_1} &=& g^{y'}{}_{x_1}\biggl( -\,\delta^{y_0}{}_{y'} 
+\,\sum\limits_{k=2}^\infty\,{\frac {1}{k!}}\,\alpha^{y_0}{}_{y'y_2\!\dots \!y_{k+1}}\sigma^{y_2}\cdots\sigma^{y_{k+1}}\biggr)\!,\label{app_expansion_1}\\
\sigma^{y_0}{}_{y_1} &=& \delta^{y_0}{}_{y_1} 
-\,\sum\limits_{k=2}^\infty\,{\frac {1}{k!}}\,\beta^{y_0}{}_{y_1y_2\dots y_{k+1}} \sigma^{y_2}\!\cdots\!\sigma^{y_{k+1}}, \label{app_expansion_2} \\
g^{y_0}{}_{x_1 ; x_2} &=& g^{y'}{\!}_{x_1} g^{y''}{\!}_{x_2}\biggl({\frac 12} 
R^{y_0}{}_{y'y''y_3}\sigma^{y_3}\nonumber\\ 
&&\qquad\qquad \!+\!\sum\limits_{k=2}^\infty\,{\frac {1}{k!}}\,\gamma^{y_0}{}_{y'y''y_3\dots y_{k+2}}\sigma^{y_3}\!\cdots\!\sigma^{y_{k+2}}\!\biggr)\!,\label{app_expansion_3} \\
g^{y_0}{}_{x_1 ; y_2} &=& g^{y'}{\!}_{x_1} \biggl({\frac 12} R^{y_0}{}_{y'y_2y_3}\sigma^{y_3}\nonumber\\ 
&&\qquad\qquad \!+\!\sum\limits_{k=2}^\infty\,{\frac {1}{k!}}\,\gamma^{y_0}{}_{y'y_2y_3\dots y_{k+2}}\sigma^{y_3}\!\cdots\!\sigma^{y_{k+2}}\!\biggr).\label{app_expansion_4}
\end{eqnarray}
The coefficients $\alpha, \beta, \gamma$ in these expansions are polynomials constructed from the Riemann curvature tensor and its covariant derivatives. The first coefficients read (as one can also check using computer algebra \cite{Ottewill:2011}):
\begin{eqnarray}
\alpha^{y_0}{}_{y_1y_2y_3} &=& -\,\frac{1}{3} R^{y_0}{}_{(y_2y_3)y_1},\label{a1}\\
\beta^{y_0}{}_{y_1y_2y_3} &=& \frac{2}{3}R^{y_0}{}_{(y_2y_3)y_1},\label{be1}\\
\alpha^{y_0}{}_{y_1y_2y_3y_4} &=& \frac{1}{2} \nabla_{(y_2}R^{y_0}{}_{y_3y_4)y_1},\label{al2}\\
\beta^{y_0}{}_{y_1y_2y_3y_4} &=& -\,\frac{1}{2} \nabla_{(y_2} R^{y_0}{}_{y_3y_4)y_1},\label{be2}\\
\alpha^{y_0}{}_{y_1y_2y_3y_4y_5}  &=& -\,\frac{7}{15} R^{y_0}{}_{(y_2y_3|y'|} R^{y'}{}_{y_4y_5)y_1} 
-\,\frac{3}{5} \nabla_{(y_5} \nabla_{y_4} R^{y_0}{}_{y_2y_3)y_1},\label{al3}  \\
\beta^{y_0}{}_{y_1y_2y_3y_4y_5} &=& \frac{8}{15} R^{y_0}{}_{(y_2y_3|y'|} R^{y'}{}_{y_4y_5)y_1} 
+\,\frac{2}{5}  \nabla_{(y_5} \nabla_{y_4} R^{y_0}{}_{y_2y_3)y_1} ,\label{be3}\\
\gamma^{y_0}{}_{y_1y_2y_3y_4}&=& \frac{1}{3} \nabla_{(y_3} R^{y_0}{}_{|y_1|y_4)y_2}, \label{ga} \\
\gamma^{y_0}{}_{y_1y_2y_3y_4y_5} &=& \frac{1}{4}  R^{y_0}{}_{y_1y'(y_3} R^{y'}{}_{y_4y_5)y_2} 
+\,\frac{1}{4} \nabla_{(y_5}\nabla_{y_4} R^{y_0}{}_{|y_1y_2|y_3)}. \label{ga2}
\end{eqnarray}
These results allow us to derive the third derivatives of the world function appearing in (\ref{eta_2nd_deriv_alternative_1}) and (\ref{eta_2nd_deriv_alternative_2}), i.e.\ we have up to the second order in the deviation vector:
\begin{eqnarray}
\sigma^{y_0}{}_{y_1 y_2} &=& - \frac{2}{3} R^{y_0}{}_{(y_2 y_3) y_1} \sigma^{y_3} - \frac{1}{2} \left( \frac{1}{2}  \nabla_{y_2} R^{y_0}{}_{(y_3 y_4) y_1} \right. \nonumber \\
&& \left. - \frac{1}{3} \nabla_{y_3} R^{y_0}{}_{(y_2 y_4) y_1} \right)   \sigma^{y_3} \sigma^{y_4}\nonumber\\ 
&& -\,{\frac 16}\lambda^{y_0}{}_{y_1 y_2 y_3y_4y_5}\sigma^{y_3}\sigma^{y_4}\sigma^{y_5}  + {\mathcal O}(\sigma^4), \label{sigy123} \\
%---
\sigma^{y_0}{}_{y_1 x_2} &=& g^{y'}{}_{x_2} \left( \frac{2}{3}  R^{y_0}{}_{(y' y_3) y_1} \sigma^{y_3}
-\,\frac{1}{4} \nabla_{(y'} R^{y_0}{}_{y_3 y_4) y_1} \sigma^{y_3} \sigma^{y_4} \right. \nonumber\\ 
&& \left. + \,{\frac 16}\mu^{y_0}{}_{y_1 y' y_3y_4y_5}\sigma^{y_3}\sigma^{y_4}\sigma^{y_5} \right) + {\mathcal O}(\sigma^4), \\
%---
\sigma^{y_0}{}_{x_1 x_2} &=& - g^{y'}{}_{x_1} g^{y''}{}_{x_2} \left[\left(\frac{1}{2} R^{y_0}{}_{y' y'' y_3} - \frac{1}{3} R^{y_0}{}_{( y'' y_3) y'} \right) \sigma^{y_3} \right. \nonumber \\
&& +\left( \frac{1}{6}  \nabla_{(y_3} R^{y_0}{}_{| y' | y_4 ) y''}  + \frac{1}{4} \nabla_{(y''} R^{y_0}{}_{y_3 y_4) y'} \right) \sigma^{y_3} \sigma^{y_4} \nonumber\\
&& \left. +\,{\frac 16}\nu^{y_0}{}_{y' y'' y_3y_4y_5}\sigma^{y_3}\sigma^{y_4}\sigma^{y_5} \right]  + {\mathcal O}(\sigma^4). \label{sigx123}
\end{eqnarray}
Here we introduced a compact notation for the combinations of the second covariant derivatives of the curvature and the quadratic polynomial of the curvature tensor (in symbolic form, ``$\nabla\nabla R + R\cdot R$''):
\begin{eqnarray}
\lambda^{y_0}{}_{y_1 y_2 y_3y_4y_5} \!\!&=&\!\! \beta^{y_0}{}_{y_1 y_3 y_4y_5;y_2} + \beta^{y_0}{}_{y_1 y_2 y_3y_4y_5}
-\,3\beta^{y_0}{}_{y_1 y' (y_3}\beta^{y'}{}_{|y_2|y_4y_5)},\label{lambda}\\
%---
\mu^{y_0}{}_{y_1 y_2 y_3y_4y_5} \!\!&=&\!\! \beta^{y_0}{}_{y_1 y_2 y_3y_4y_5} 
-\,3\beta^{y_0}{}_{y_1 y' (y_3}\alpha^{y'}{}_{|y_2|y_4y_5)},\label{mu}\\
%---
\nu^{y_0}{}_{y_1 y_2 y_3y_4y_5} \!\!&=&\!\! \gamma^{y_0}{}_{y_1 y_2y_3 y_4y_5} + \alpha^{y_0}{}_{y_1 y_2 y_3y_4y_5}
-\,3\alpha^{y_0}{}_{y_1 y' (y_3}\alpha^{y'}{}_{|y_2|y_4y_5)}\nonumber  \\
\!\!&&\!\! -\,\frac{1}{4} R^{y'}{}_{y_1 y_2 (y_3}\alpha^{y_0}{}_{|y'|y_4 y_5)}.\label{nu}
\end{eqnarray}
Substituting the coefficients of the expansions (\ref{app_expansion_1})-(\ref{app_expansion_3}) we obtain the explicit (complicated) expressions which we do not display here.

For the symmetrized versions of (\ref{sigy123}) and (\ref{sigx123}) we obtain 
\begin{eqnarray}
\sigma^{y_0}{}_{(y_1 y_2)} &=& \frac{1}{3} R^{y_0}{}_{(y_1 y_2) y_3} \sigma^{y_3}\nonumber \\
&& -\, \frac{1}{4} \left( \nabla_{(y_1} R^{y_0}{}_{|y_3 y_4 | y_2)} 
+ \frac{1}{3} \nabla_{y_3} R^{y_0}{}_{(y_1 y_2) y_4} \right) \sigma^{y_3} \sigma^{y_4}\nonumber\\ 
&& -\,{\frac 16}\lambda^{y_0}{}_{(y_1 y_2) y_3y_4y_5}\sigma^{y_3}\sigma^{y_4}\sigma^{y_5} + {\mathcal O}(\sigma^4), \\
%---
\sigma^{y_0}{}_{(x_1 x_2)} &=& g^{y'}{}_{(x_1} g^{y''}{}_{x_2)} \left[-\frac{2}{3} R^{y_0}{}_{(y' y'') y_3} \sigma^{y_3} \right. \nonumber \\
&& + \,\frac{1}{4} \left(  \nabla_{y_3} R^{y_0}{}_{( y' y'' ) y_4 } 
-\,\frac{1}{3} \nabla_{(y'} R^{y_0}{}_{|y_3 y_4 | y'')} \right) \sigma^{y_3} \sigma^{y_4}  \nonumber \\ 
&&\left. -\,{\frac 16}\nu^{y_0}{}_{(y' y'') y_3y_4y_5}\sigma^{y_3}\sigma^{y_4}\sigma^{y_5}\right] + {\mathcal O}(\sigma^4),
\end{eqnarray}
Furthermore we need the expansions of $K^{x}{}_{y}$ and $\stackrel{-1}{\sigma}{}\!\!^{x}{}_y = -\,H^{x}{}_{y}$: 
\begin{eqnarray}
\stackrel{-1}{\sigma}{}\!\!^{x_1}{}_{y_2} &=& - \,g^{x_1}{}_{y'} \left( \delta^{y'}{}_{y_2} - \frac{1}{6} R^{y'}{}_{(y_3 y_4) y_2} \sigma^{y_3} \sigma^{y_4}  \right. \nonumber \\
&& \left. + \,\frac{1}{12} \nabla_{(y_3} R^{y'}{}_{y_4 y_5)y_2}  \sigma^{y_3} \sigma^{y_4} \sigma^{y_5} \right) + {\mathcal O}(\sigma^4) ,\\
%---
 K^{x_1}{}_{y_2} &=& g^{x_1}{}_{y'} \left( \delta^{y'}{}_{y_2} - \frac{1}{2} R^{y'}{}_{(y_3 y_4) y_2} \sigma^{y_3} \sigma^{y_4} \right. \nonumber \\
&& \left. +\,\frac{1}{6} \nabla_{(y_3} R^{y'}{}_{y_4 y_5)y_2}  \sigma^{y_3} \sigma^{y_4} \sigma^{y_5} \right) + {\mathcal O}(\sigma^4).
\end{eqnarray}
From this one can derive the recurring term in (\ref{eta_2nd_deriv_alternative_2}) up to the needed order, i.e.
\begin{eqnarray}
&& \left( K^{x_1}{}_{y_2} u^{y_2} - H^{x_1}{}_{y_2} \frac{D \sigma^{y_2}}{dt} \right) 
=  g^{x_1}{}_{y'} \Biggl[ u^{y'} - \frac{D \sigma^{y'}}{dt}\nonumber \\
&& - \,\frac{1}{2} R^{y'}{}_{(y_3 y_4) y_2} \sigma^{y_3} \sigma^{y_4} \left( u^{y_2} - \frac{1}{3}\frac{D \sigma^{y_2}}{dt}\right) \nonumber\\
&& +\,\frac{1}{6}\nabla_{(y_3} R^{y'}{}_{y_4 y_5)y_2} u^{y_2}\sigma^{y_3} \sigma^{y_4} \sigma^{y_5}\Biggr]  + {\mathcal O}(\sigma^4).
\end{eqnarray}
With these expansions at hand we are finally able to develop the deviation equation (\ref{eta_2nd_deriv_alternative_2}) up to the third order.

Denote $\tilde{a}^{y_1} = g^{y_1}{}_{x_2} \tilde{a}^{x_2}$ in accordance with the definition of the parallel propagator, and introduce 
\begin{eqnarray}
\phi^{y_1}{}_{y_2 y_3 y_4y_5y_6} = \lambda^{y_1}{}_{y_2 y_3 y_4y_5y_6} - 2\mu^{y_1}{}_{y_2 y_3 y_4y_5y_6}
+ \nu^{y_1}{}_{y_2 y_3 y_4y_5y_6}. \label{phi}
\end{eqnarray}
The deviation equation up to the third order reads
\begin{eqnarray}
\frac{D^2}{dt^2} \eta^{y_1} &=& \tilde{a}^{y_1} \left(\frac{d\tilde{t}}{dt} \right)^2 - a^{y_1} +  \frac{d t}{d \tilde{t}} \frac{d^2\tilde{t}}{dt^2} u^{y_1} + \frac{D \eta^{y_1}}{dt} \frac{d t}{d \tilde{t}} \frac{d^2\tilde{t}}{dt^2}\nonumber \\
&& - \eta^{y_4} R^{y_1}{}_{y_2 y_3 y_4} \left(  u^{y_2} u^{y_3} + 2 u^{y_3} \frac{D \eta^{y_2}}{dt} \right) 
\nonumber \\
&&  + \eta^{y_4} \eta^{y_5} \Bigg\{ u^{y_2} u^{y_3}  \left( \frac{1}{2} \nabla_{y_2} R^{y_1}{}_{y_4 y_5 y_3 } - \frac{1}{3}  \nabla_{y_4} R^{y_1}{}_{y_2 y_3 y_5} \right) \nonumber \\
&&+  \frac{1}{3} R^{y_1}{}_{y_4 y_5 y_2} \left[ a^{y_2} + \frac{1}{2} \tilde{a}^{y_2} \left(\frac{d\tilde{t}}{dt} \right)^2 - u^{y_2} \frac{d t}{d \tilde{t}} \frac{d^2\tilde{t}}{dt^2} \right]\Bigg\} \nonumber \\
&&  - {\frac 16}\eta^{y_4} \eta^{y_5} \eta^{y_6} \Bigg\{ \phi^{y_1}{}_{y_2 y_3 y_4y_5y_6} u^{y_2}u^{y_3} \nonumber \\ 
&&- {\frac 12}\nabla_{(y_4}R^{y_1}{}_{y_5y_6)y_2} \left[ a^{y_2} + \tilde{a}^{y_2} \left(\frac{d\tilde{t}}{dt} \right)^2 - u^{y_2} \frac{d t}{d \tilde{t}} \frac{d^2\tilde{t}}{dt^2} \right]\Bigg\}\nonumber \\
&& - {\frac 12}u^{y'}\frac{D \eta^{y''}}{dt}\eta^{y_2}\eta^{y_3}\Bigg( - \nabla_{(y''} R^{y_1}{}_{y_2 y_3) y_1 } + \nabla_{y_2} R^{y_1}{}_{(y' y'') y_3} \nonumber \\
&& - {\frac 13}\nabla_{(y'} R^{y_1}{}_{|y_2 y_3|y'')} \Bigg) - {\frac 23}\frac{D \eta^{y_2}}{dt}\frac{D \eta^{y_3}}{dt}\eta^{y_4}\,R^{y_1}{}_{y_2y_3y_4} + \mathcal{O}(\sigma^4).\nonumber\\ \label{eta_2nd_deriv_compact_2}
\end{eqnarray}
We would like to stress that the generalized deviation equation (\ref{eta_2nd_deriv_compact_2}) is completely general. In particular, it allows for a comparison of two general, i.e.\ not necessarily geodetic, world lines in spacetime. Various special cases of (\ref{eta_2nd_deriv_compact_2}) qualitatively reproduce all the previous results in the literature, see in particular \cite{Hodgkinson:1972,Bazanski:1977:1,Aleksandrov:Piragas:1978,Schutz:1985,Chicone:Mashhoon:2002,Mullari:Tammelo:2006,Vines:2014}.

\subsection{Method 2: Reference frame (inertial and gravitational effects) }\label{subsec_theo_found_2}

The above discussion of the deviation equation made clear that a suitable choice of coordinates is crucial for the successful determination of the gravitational field. In particular, the operational realization of the coordinates is of importance when it comes to actual measurements. 

From an experimentalists perspective so-called (generalized) Fermi coordinates appear to be realizable operationally. There have been several suggestions for such coordinates in the literature in different contexts \cite{Fermi:1922:1:2:3,Fermi:1962:US,Veblen:1922,Veblen:Thomas:1923,Synge:1931,Walker:1932,Synge:1960,Manasse:Misner:1963,MTW:1973,Ni:1977,Mashhoon:1977,Ni:Zimmermann:1978,Li:Ni:1978,Ni:1978,Li:Ni:1979,Li:Ni:1979:1,Ashby:etal:1986,Eisele:1987,Fukushima:1988,Semerak:1993,Marzlin:1994:1,Bini:etal:2005,Chicone:Mashhoon:2006:1,Klein:Collas:2008:1,Klein:Collas:2010:2,Delva:etal:2012,Turyshev:etal:2012}. In the following we are going to derive the line element in the vicinity of a world line, representing an observer in an arbitrary state of motion, in generalized Fermi coordinates.

\paragraph{Fermi normal coordinates}\label{fermi_subpara}

Following \cite{Veblen:Thomas:1923} we start by taking successive derivatives of the usual geodesic equation. This generates a set of equations of the form (for $n \geq 2$) 
\begin{eqnarray}
\frac{d^n x^a}{ds^n} &=& - \Gamma_{b_1 \dots b_{n}}{}^a \, \frac{d x^{b_1}}{ds} \cdots \frac{d x^{b_{n}}}{ds},
\label{geoddiffset}
\end{eqnarray}
where the $\Gamma$ objects with $n \geq 3$ lower indices are defined by the recurrent relation
\begin{eqnarray}
\Gamma_{b_{1}\dots b_{n}}{}^a := \partial_{(b_1} \Gamma_{b_2 \dots b_{n})}{}^a - (n-1)\, \Gamma_{c (b_1 \dots b_{n-2}}{}^{a} \, \Gamma_{b_{n-1} b_n)}{}^{c} \label{gammadefinition}
\end{eqnarray}
from the components of the linear connection $\Gamma_{bc}{}^a$. 
A solution $x^a=x^a(s)$ of the geodesic equation may then be expressed as a series
\begin{eqnarray}
x^a &=& \left. x^a  \right|_{0} + s \left. \frac{dx^a}{ds} \right|_{0} + \frac{s^2}{2} \left. \frac{d^2x^a}{ds^2} \right|_{0} +  \frac{s^3}{6} \left. \frac{d^3x^a}{ds^3} \right|_{0} + \cdots \nonumber \\
&=& q^a + s v^a - \frac{s^2}{2} \stackrel{0}{\Gamma}_{bc}{\!}^a \, v^b v^c - \frac{s^3}{6} \stackrel{0}{\Gamma}_{bcd}{\!}^a \, v^b v^c v^d - \cdots\,,
\end{eqnarray}
where in the last line we used $q^a :=\left. x^a  \right|_{0}$, $v^a := \left. \frac{dx^a}{ds} \right|_{0}$, and $\stackrel{0}{\Gamma}_{\dots}{}^a:=\left. \Gamma_{\dots}{}^a \right|_{0}$ for constant quantities at the point around which the series development is performed.

\begin{figure}
\begin{center}
\includegraphics[width=7.8cm,angle=-90]{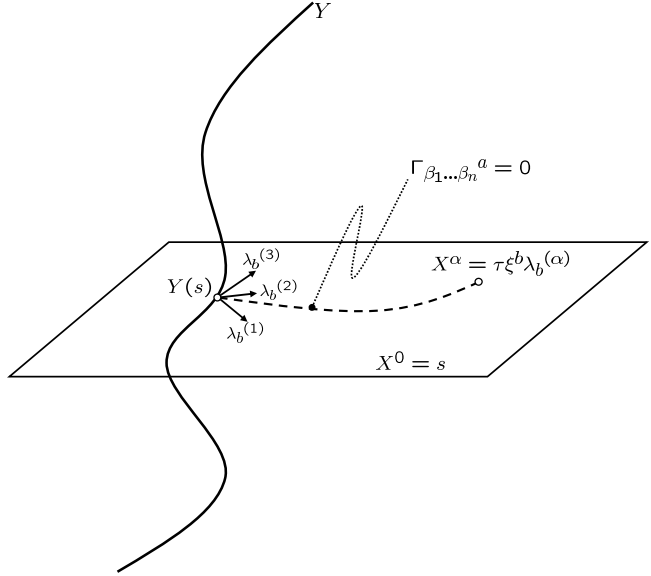}
\end{center}
\caption{\label{fig_1} Construction of the coordinate system around the reference curve $Y$. Coordinates of a point $X$ in the vicinity of $Y(s)$ -- $s$ representing the proper time along $Y$ -- are constructed by means of a tetrad $\lambda_b{}^{(\alpha)}$. Here $\tau$ is the proper time along the (spacelike) geodesic connecting $Y$ and $X$. By choosing a linear ansatz for the coordinates the derivatives of the connection vanish along the geodesic connecting $Y$ and $X$.}
\end{figure}

Now let us setup coordinates centered on the reference curve $Y$ to describe an adjacent point $X$. For this we consider a unique geodesic connecting $Y$ and $X$. We define our coordinates in the vicinity of a point on $Y(s)$, with proper time $s$, by using a tetrad $\lambda_b{}^{(\alpha)}$ which is Fermi transported along $Y$, i.e.\ 
\begin{eqnarray}
X^0 = s, \quad \quad X^\alpha = \tau \xi^b \lambda_b{}^{(\alpha)}. \label{fermi_ansatz}
\end{eqnarray}
Here $\alpha=1,\dots,3$, and $\tau$ is the proper time along the (spacelike) geodesic connecting $Y(s)$ and $X$. The $\xi^b$ are constants, and it is important to notice that the tetrads are functions of the proper time $s$ along the reference curve $Y$, but independent of $\tau$. See figure \ref{fig_1} for further explanations. By means of this linear ansatz (\ref{fermi_ansatz}) for the coordinates in the vicinity of $Y$, we obtain for the derivatives w.r.t.\ $\tau$ along the connecting geodesic ($n \geq 1$):
\begin{eqnarray}
\frac{d^n X^0}{d \tau^n} = 0,\qquad 
\frac{d X^\alpha}{d \tau} = \xi^b \lambda_b{}^{(\alpha)}, \quad \frac{d^{n+1} X^\alpha}{d \tau^{n+1}} =0 .
\end{eqnarray}
In other words, in the chosen coordinates (\ref{fermi_ansatz}), along the geodesic connecting $Y$ and $X$, one obtains for the derivatives ($n \geq 2$)
\begin{eqnarray}
\Gamma_{b_1 \dots b_{n}}{}^a \, \frac{d X^{b_1}}{d\tau} \cdots \frac{d X^{b_{n}}}{d\tau}=0.
\label{geoddiffset_2}
\end{eqnarray}
This immediately yields 
\begin{eqnarray}
\Gamma_{\beta_1 \dots \beta_{n}}{}^a = 0, \label{connection_deriv_condition}
\end{eqnarray}
along the connecting curve, in the region covered by the linear coordinates as defined above.

The Fermi normal coordinate system cannot cover the whole spacetime manifold. By construction, it is a good way to describe the physical phenomena in a small region around the world line of an observer. The smallness of the corresponding domain depends on the motion of the latter, in particular, on the magnitudes of acceleration $|a|$ and angular velocity $|\omega|$ of the observer which set the two characteristic lengths: $\ell_{\textrm{tr}} = c^2/|a|$ and $\ell_{\textrm{rot}} = c/|\omega|$. The Fermi coordinate system $X^\alpha$ provides a good description for the region $|X|/\ell \ll 1$. For example, this condition is with a high accuracy valid in terrestrial laboratories since $\ell_{\textrm{tr}} = c^2/|g_\oplus| \approx 10^{16}$m (one light year), and $\ell_{\textrm{rot}} = c/|\Omega_\oplus| \approx 4\times 10^{12}$m (27 astronomical units). Note, however, that for a particle accelerated in a storage ring $\ell \approx 10^{-6}$m. Furthermore, the region of validity of the Fermi coordinate system is restricted by the strength of the gravitational field in the region close to the reference curve, $\ell_{\textrm{grav}} = \textrm{min}\{|R_{abcd}|^{-1/2}, |R_{abcd}|/|R_{abcd,e}|\}$, so that the curvature should have not yet caused geodesics to cross. We always assume that there is a unique geodesic connecting $Y$ and $X$.

\paragraph{Explicit form of the connection}

At the lowest order, in flat spacetime, the connection of a noninertial system that is accelerating with $a^\alpha$ and rotating with angular velocity $\omega^\alpha$ at the origin of the coordinate system is
\begin{eqnarray}
\Gamma_{00}{}^{0} = \Gamma_{\alpha \beta}{}^{c}=0,\quad
\Gamma_{00}{}^{\alpha} = a^\alpha,\quad 
\Gamma_{0\alpha}{}^{0} = a_\alpha, \quad
\Gamma_{0\beta}{}^{\alpha} = - \varepsilon^{\alpha}{}_{\beta \gamma} \omega^\gamma. \label{connection_expl}
\end{eqnarray}
Hereafter $\varepsilon_{\alpha\beta\gamma}$ is the 3-dimensional totally antisymmetric Levi-Civita symbol, and the Euclidean 3-dimensional metric $\delta_{\alpha\beta}$ is used to raise and lower the spatial (Greek) indices, in particular $a_\alpha = \delta_{\alpha\beta}a^\beta$ and $\varepsilon^{\alpha}{}_{\beta \gamma} = \delta^{\alpha\delta}\varepsilon_{\delta\beta \gamma}$.
For the time derivatives we have
\begin{eqnarray}
\partial_{0}\Gamma_{00}{}^{0}&=&\partial_{0} \Gamma_{\alpha \beta}{}^{c}=0,\quad
\partial_{0} \Gamma_{00}{}^{\alpha} = \partial_{0} a^\alpha =: b^{\alpha},\quad \nonumber \\
\partial_{0} \Gamma_{0\alpha}{}^{0} &=& b_{\alpha}, \,
\partial_{0} \Gamma_{0\beta}{}^{\alpha} = - \varepsilon^{\alpha}{}_{\beta \gamma} \partial_{0} \omega^\gamma =: - \varepsilon^{\alpha}{}_{\beta \gamma} \eta^\gamma. \label{0_connection_expl}
\end{eqnarray}
From the definition of the curvature we can express the next order of derivatives of the connection in terms of the curvature:
\begin{eqnarray}
\partial_{\alpha}\Gamma_{00}{}^{0} &=& b_\alpha - a_\beta \varepsilon^{\beta}{}_{\alpha \gamma} \omega^\gamma, \nonumber\\
\partial_{\alpha}\Gamma_{00}{}^{\beta}&=& -\,R_{0 \alpha 0}{}^\beta-\varepsilon^\beta{}_{\alpha \gamma} \eta^{\gamma} + a_\alpha a^\beta - \delta_\alpha^\beta \omega_\gamma \omega^\gamma + \omega_\alpha \omega^\beta, \nonumber \\
\partial_{\alpha}\Gamma_{0\beta}{}^{0}&=& -\,R_{0 \alpha \beta}{}^0 - a_\alpha a_\beta , \nonumber \\
\partial_{\alpha}\Gamma_{0\beta}{}^{\gamma}&=& -\,R_{0 \alpha \beta}{}^\gamma + \varepsilon^\gamma{}_{\alpha \delta} \omega^\delta a_\beta. \label{1_connection_expl}
\end{eqnarray}

Using (\ref{connection_deriv_condition}), we derive the spatial derivatives 
\begin{eqnarray}
\partial_{\alpha}\Gamma_{\beta \gamma}{}^{d}&=& \frac{2}{3} R_{\alpha (\beta \gamma)}{}^d, \label{1_connection_expl_1} 
\end{eqnarray}
see also the general solution given in the appendix B of \cite{Puetzfeld:Obukhov:2016:1}.

\paragraph{Explicit form of the metric}

In order to determine, in the vicinity of the reference curve $Y$, the form of the metric at the point $X$ in coordinates $y^a$ centered on $Y$, we start again with an expansion of the metric around the reference curve
\begin{eqnarray}
\left. g_{ab} \right|_{X} &=& \left. g_{ab}  \right|_{Y} + \left. g_{ab,c} \right|_{Y} y^c + \frac{1}{2} \left. g_{ab,cd} \right|_{Y} y^c y^d  + \cdots\,. 
\end{eqnarray}
Of course in normal coordinates we have $\left. g_{ab}  \right|_{Y} = {\eta}_{ab}$, whereas the derivatives of the metric have to be calculated, and the result actually depends on which type of coordinates we want to use. The derivatives of the metric may be expressed just by successive differentiation of the metricity condition $\nabla_c g_{ab} =0$:
\begin{eqnarray}
g_{ab,c} &=& 2 \, g_{d(a} \Gamma_{b)c}{}^d, \nonumber \\
g_{ab,cd} &=& 2 \, \left( \partial_d g_{e(a} \Gamma_{b)c}{}^e + \partial_d \Gamma_{c(a}{}^e  g_{b)e} \right), \nonumber \\
&\vdots&  \label{metricity_cond_explicit}
\end{eqnarray}
In other words, we can iteratively determine the metric by plugging in the explicit form of the connection and its derivatives from above.

In combination with (\ref{connection_expl}) one finds:
\begin{eqnarray}
g_{00,0} &=& g_{0\alpha,0} = g_{\alpha \beta,0} = g_{\alpha \beta, \gamma}  = 0, \nonumber \\
g_{00,\alpha} &=& 2 a_\alpha, \quad g_{0 \alpha , \beta} = \varepsilon_{\alpha \beta \gamma} \omega^\gamma. \label{1_metric_explicit}
\end{eqnarray}

For the second order derivatives of metric we obtain, again using (\ref{metricity_cond_explicit}) in combination with (\ref{0_connection_expl}), (\ref{1_connection_expl}), and (\ref{1_metric_explicit}):
\begin{eqnarray}
g_{00,00} &=& g_{0\alpha,0 0} = g_{\alpha \beta,00}= g_{\alpha \beta,\gamma 0} = 0, \quad g_{00,\alpha 0} = 2 b_\alpha, \nonumber \\
g_{0\alpha,\beta 0} &=& - \varepsilon^{\gamma}{}_{\beta \delta} \eta^{\delta} g_{\alpha \gamma} = \varepsilon_{\alpha \beta \gamma} \eta^{\gamma}, \nonumber \\
g_{00,\alpha \beta} &=& -\,2 R_{0\beta \alpha}{}^0 + 2 a_\alpha a_\beta - 2 \delta_{\alpha \beta} \omega_\gamma \omega^\gamma + 2 \omega_\alpha \omega_\beta ,\nonumber \\
g_{0\alpha,\beta \gamma} &=&  -\,{\frac 43} R_{\alpha(\beta\gamma)}{}^0,\quad
g_{\alpha \beta, \gamma \delta} = {\frac 23} R_{\gamma (\alpha \beta) \delta}. \label{2_metric_explicit}
\end{eqnarray}
Note that $R_{0\beta \alpha}{}^0 + R_{\alpha 0\beta}{}^0 + R_{\beta \alpha 0}{}^0 \equiv 0$, in view of the Ricci identity. Since $R_{\beta \alpha 0}{}^0 = 0$, we thus find $R_{0\beta \alpha}{}^0 = R_{0(\beta \alpha)}{}^0$. 

As a result, we derive the line element in the Fermi coordinates (up to the second order):
\begin{eqnarray}
\left. ds^2 \right|_X(y^{0},y^{\alpha}) &=& (dy^{0})^2 \bigr[1 + 2 a_{\alpha} y^{\alpha} + 2 b_\alpha y^\alpha y^0 \nonumber \\
&&+ (a_\alpha a_\beta - \delta_{\alpha \beta} \omega_\gamma \omega^\gamma + \omega_\alpha \omega_\beta - R_{0\alpha\beta 0}) y^\alpha y^\beta \bigr]    \nonumber \\ 
&&+ 2 dy^{0} dy^{\alpha} \bigr[ \varepsilon_{\alpha \beta \gamma} \omega^\gamma y^\beta + \varepsilon_{\alpha \beta \gamma} \eta^\gamma y^\beta y^0 - \frac{2}{3} R_{\alpha \beta \gamma 0} y^{\beta} y^{\gamma} \bigr] \nonumber \\
&&-dy^{\alpha} dy^{\beta}\bigr[\delta_{\alpha\beta} - \frac{1}{3} R_{\gamma \alpha \beta \delta} y^{\gamma} y^{\delta} \bigr] + {\mathcal O}(3).
\label{fermi_normal_ds}
\end{eqnarray}
It is worthwhile to notice that we can recast this result into
\begin{eqnarray}
\left. ds^2 \right|_X(y^{0},y^{\alpha}) &=& (dy^{0})^2 \bigr[1 + 2 \overline{a}_{\alpha} y^{\alpha} + (\overline{a}_\alpha \overline{a}_\beta - \delta_{\alpha \beta} \overline{\omega}_\gamma \overline{\omega}^\gamma + \overline{\omega}_\alpha \overline{\omega}_\beta  \nonumber \\ 
&& - R_{0\alpha\beta 0}) y^\alpha y^\beta \bigr]+ 2 dy^{0} dy^{\alpha} \bigr[ \varepsilon_{\alpha \beta \gamma} \overline{\omega}^\gamma y^\beta - \frac{2}{3} R_{\alpha \beta \gamma 0} y^{\beta} y^{\gamma} \bigr]     \nonumber \\ 
&&-dy^{\alpha} dy^{\beta}\bigr[\delta_{\alpha\beta} - \frac{1}{3} R_{\gamma \alpha \beta \delta} y^{\gamma} y^{\delta} \bigr] + {\mathcal O}(3),
\label{fermi_normal_ds1}
\end{eqnarray}
by introducing $\overline{a}_{\alpha} = a_\alpha + y^0\partial_0a_\alpha = a_\alpha + y^0 b_\alpha$ and $\overline{\omega}_{\alpha} = \omega_\alpha + y^0\partial_0\omega_\alpha = \omega_\alpha + y^0 \eta_\alpha$ which represent the power expansion of the time dependent acceleration and angular velocity.

\subsection{Method 2: Apparent behavior of clocks}\label{behavior_clocks_sec}\label{subsec_theo_found_3}

The results from the last section may now be used to describe the behavior of clocks in the vicinity of the reference world line, around which the coordinates were constructed. 

There is one interesting peculiarity about writing the metric like in (\ref{fermi_normal_ds}), i.e.\ one obtains clock effects which depend on the acceleration of the clock (just integrate along a curve in those coordinates and the terms with $a$ and $\omega$ will of course contribute to the proper time along the curve). This behavior of clocks is of course due to the choice of the noninertial observer, and they are {\it only} present along curves which do not coincide with the observers world line. Recall that, by construction, one has Minkowski's metric along the world line of the observer, which is also the center of the coordinate system in which (\ref{fermi_normal_ds}) is written -- all inertial effects vanish at the origin of the coordinate system. 

\paragraph{Flat case}\label{paragraph_fermi_normal_flat_freq_ratio}

We start with the flat spacetime and switch to a quantity which is directly measurable, i.e.\ the proper time quotient of two clocks located at $Y$ and $X$. It is worthwhile to note that for a flat spacetime, $R_{ijk}{}^l = 0$, the interval (\ref{fermi_normal_ds}) reduces to the Hehl-Ni \cite{Hehl:Ni:1990} line element of a noninertial (rotating and accelerating) system:
\begin{eqnarray}
\left.ds^2 \right|_X(y^{0},y^{\alpha}) &=& (1 + \overline{a}_\alpha y^\alpha)^2(dy^{0})^2 - \delta_{\alpha\beta} (dy^{\alpha} + \varepsilon^\alpha{}_{\mu\nu}\overline{\omega}^\mu y^\nu\,dy^0)\nonumber \\ 
&&\times(dy^{\beta} + \varepsilon^\beta{}_{\rho\sigma}\overline{\omega}^\rho y^\sigma\,dy^0) + {\mathcal O}(3),
\label{fermi_normal_ds2}
\end{eqnarray}
From (\ref{fermi_normal_ds}) we derive
\begin{eqnarray}
\left(\frac{ds|_X}{ds|_Y}\right)^2 &=&\left(\frac{dy^{0}}{ds|_Y}\right)^2 \bigr[ 1 - \delta_{\alpha \beta} v^\alpha v^\beta + 2 a_\alpha y^\alpha + 2 b_\alpha y^\alpha y^0 \nonumber \\
&&+ y^\alpha y^\beta \left( a_\alpha a_\beta - \delta_{\alpha \beta} \omega_\gamma \omega^\gamma + \omega_\alpha \omega_\beta \right) \nonumber \\
&&+ 2 v^\alpha \varepsilon_{\alpha \beta \gamma}  \left(y^\beta \omega^\gamma + y^0 y^\beta \eta^\gamma \right) \bigr] + {\mathcal O}(3) \\
&=&1+ \frac{1}{1 - \delta_{\alpha \beta} v^\alpha v^\beta } \bigr[ 2 a_\alpha y^\alpha + 2 b_\alpha y^\alpha y^0  \nonumber \\
&&+  y^\alpha y^\beta \left( a_\alpha a_\beta - \delta_{\alpha \beta} \omega_\gamma \omega^\gamma + \omega_\alpha \omega_\beta \right) \nonumber \\
&&+ 2 v^\alpha \varepsilon_{\alpha \beta \gamma}  \left(y^\beta \omega^\gamma + y^0 y^\beta \eta^\gamma \right) \bigr] + {\mathcal O}(3). \label{fermi_normal_flat_freq_ratio}
\end{eqnarray}
Here we introduced the velocity $v^\alpha = dy^\alpha/dy^0$. Defining 
\begin{equation}
V^\alpha := v^\alpha + \varepsilon^\alpha{}_{\beta \gamma}\overline{\omega}^\beta y^\gamma,\label{Va}
\end{equation}
we can rewrite the above relation more elegantly as
\begin{eqnarray}
\left(\frac{ds|_X}{ds|_Y}\right)^2 &=&\left(\frac{dy^{0}}{ds|_Y}\right)^2 \bigr[ (1 + \overline{a}_\alpha y^\alpha)^2 - \delta_{\alpha \beta} V^\alpha V^\beta \bigr] + {\mathcal O}(3).\label{fermi_normal_flat_freq_ratio1}
\end{eqnarray}

Equation (\ref{fermi_normal_flat_freq_ratio}) is reminiscent of the situation which we encountered in case of the gravitational compass, i.e.\ we may look at this measurable quantity depending on how we prepare the 
\begin{eqnarray}
C\left( y^\alpha, y^0 , v^\alpha , a^\alpha, \omega^\alpha, b^\alpha, \eta^\alpha \right) := \left(\frac{ds|_X}{ds|_Y}\right)^2. \label{general_freq_ratio_definition}
\end{eqnarray}

\paragraph{Curved case}\label{paragraph_fermi_normal_curved_freq_ratio}

Now let us investigate the curved spacetime, after all we are interested in measuring the gravitational field by means of clock comparison. The frequency ratio becomes: 
\begin{eqnarray}
\left(\frac{ds|_X}{ds|_Y}\right)^2 &=&1+ \frac{1}{1 - \delta_{\alpha \beta} v^\alpha v^\beta } \bigr[ 2 a_\alpha y^\alpha + 2 b_\alpha y^\alpha y^0  \nonumber \\
&&+  y^\alpha y^\beta \left( a_\alpha a_\beta - R_{0 \alpha \beta 0} - \delta_{\alpha \beta} \omega_\gamma \omega^\gamma + \omega_\alpha \omega_\beta \right)  \nonumber \\
&& + 2 v^\alpha  \varepsilon_{\alpha \beta \gamma} \left(  y^\beta \omega^\gamma +  y^0 y^\beta \eta^\gamma \right) - \frac{4}{3} v^\alpha y^\beta y^\gamma R_{ \alpha \beta \gamma 0} \nonumber\\
&&+ \frac{1}{3} v^\alpha v^\beta y^\gamma y^\delta R_{\gamma \alpha \beta \delta} \bigr] + {\mathcal O}(3) .
\label{fermi_normal_curved_freq_ratio}
\end{eqnarray}
Analogously to the flat case in (\ref{general_freq_ratio_definition}), we introduce a shortcut for the measurable frequency ratio in a curved background, denoting its dependence on different quantities as $C\left( y^\alpha, y^0 , v^\alpha , a^\alpha, \omega^\alpha, b^\alpha, \eta^\alpha, R_{\alpha \beta \gamma \delta} \right)$.

Note that in the flat, as well as in the curved case, the frequency ratio becomes independent of $b^\alpha$ and $\eta^\alpha$ on the three-dimensional slice with fixed $y^0$ (since we can always choose our coordinate time parameter $y^0=0$), i.e.\ we have $C\left( y^\alpha, v^\alpha , a^\alpha, \omega^\alpha \right)$ and $C\left( y^\alpha, v^\alpha , a^\alpha, \omega^\alpha, R_{\alpha \beta \gamma \delta} \right)$ respectively. 

\section{Operational determination of the gravitational field}\label{sec_operational_determination}

\subsection{Method 1:  Relativistic gradiometry / Gravitational compass}\label{sec_compass}

The determination of the curvature of spacetime in the context of deviation equations has been discussed in \cite{Synge:1960,Szekeres:1965,Ciufolini:Demianski:1986}. In particular, Szekeres coined in \cite{Szekeres:1965} the notion of a ``gravitational compass.'' From now on we will adopt this notion for a set of suitably prepared test bodies which allow for the measurement of the curvature and, thereby, the gravitational field. 

The operational procedure is to monitor the accelerations of a set of test bodies w.r.t.\ to an observer moving on the reference world line $Y$. A mechanical analogue would be to measure the forces between the test bodies and the reference body via a spring connecting them. 

\paragraph{Rewriting the deviation equation}\label{sec_compass_with_standard_dev}

We now describe the configurations of test bodies which allow for a complete determination of all curvature components in a Riemannian background spacetime. For concreteness, our analysis will be based on the standard geodesic deviation equation, as well as one of its generalizations.
Our starting point is the standard geodesic deviation equation, i.e. 
\begin{eqnarray}
\frac{D^2}{ds^2}\eta^a = R^a{}_{bcd} u^b \eta^c u^d. \label{compass_start}
\end{eqnarray}
Since we want to express the curvature in terms of measured quantities, i.e.\ the velocities and the accelerations, we rewrite this equation in terms of the standard (non-covariant) derivative w.r.t.\ the proper time.

In order to simplify the resulting equation we employ normal coordinates, i.e.\ we have on the world line of the reference test body
\begin{eqnarray}
\Gamma_{ab}{}^c |_Y= 0, \quad \quad \partial_a \Gamma_{bc}{}^d|_Y = \frac{2}{3} R_{a(bc)}{}^d.
\end{eqnarray}

In terms of the standard total derivative w.r.t.\ to the proper time $s$, the deviation equation (\ref{compass_start}) takes the form:
\begin{eqnarray}
\frac{d^2}{ds^2}\eta^a &\stackrel{|_Y}{=}& \frac{2}{3} R^a{}_{bcd} u^b \eta^c u^d. \label{compass_upper}
\end{eqnarray}
However, what actually seems to be measured by a compass at the reference point $Y$ is the lower components of the relative acceleration. For the lower index position, in terms of the ordinary derivative in normal coordinates, the deviation equation (\ref{compass_start}) takes the form
\begin{eqnarray}
\frac{d^2}{ds^2}\eta_a &\stackrel{|_Y}{=}& \frac{4}{3} R_{abcd} u^b \eta^c u^d. \label{compass_lower}
\end{eqnarray} 

\paragraph{Explicit compass setup}\label{sec_compass_setup_standard_dev}

Let us consider a general 6-point compass. In addition to the reference test body on the world line we will use the following geometrical setup of the 5 remaining test bodies:
\begin{eqnarray}
&&{}^{(1)}\eta^{a}=\left(\begin{array}{c} 0 \\ 1\\ 0\\ 0\\ \end{array} \right),
{}^{(2)}\eta^{a}=\left(\begin{array}{c} 0 \\ 0\\ 1\\ 0\\ \end{array} \right),
{}^{(3)}\eta^{a}=\left(\begin{array}{c} 0 \\ 0\\ 0\\ 1\\ \end{array} \right), \nonumber \\
&&{}^{(4)}\eta^{a}=\left(\begin{array}{c} 0 \\ 1\\ 1\\ 0\\ \end{array} \right),
{}^{(5)}\eta^{a}=\left(\begin{array}{c} 0 \\ 0\\ 1\\ 1\\ \end{array} \right). \label{position_setup}
\end{eqnarray}
In addition to the positions of the compass constituents, we have to make a choice for the velocity of the reference test body / observer. In the following we will use $(m)$ different compasses, each of these compasses will have a different velocity (associated) with the reference test body. In other words, we consider $(m)$ different compasses or reference test bodies, all of which are located at the world line reference point $Y$ (at the same time), and all these $(m)$ observers measure the relative accelerations to all five test bodies placed at the positions given in (\ref{position_setup}). The left-hand sides of (\ref{compass_lower}) are the measured accelerations and in the following we refer to them by ${}^{(m,n)}A_a$. Furthermore, we also introduced the compass index ${}^{(m)}u^a$ for the velocities. In other words, for $(m)$ compasses and $(n)$ bodies in one compass, we have the following set of equations:
\begin{eqnarray}
{}^{(m,n)}A_a &\stackrel{|_Y}{=}& \frac{4}{3} R_{abcd} {}^{(m)}u^b \, {}^{(n)}\eta^c \, {}^{(m)}u^d. \label{compass_lower_setup}
\end{eqnarray}  
What remains to be chosen, apart from the $(n=1 \dots 5)$ positions of bodies in one compass, is the number $(m)$ and the actual directions in which each compass / observer shall move. Of course in the end we want to minimize both numbers, i.e.\ $(m)$ and $(n)$, which are needed to determine all curvature components.  
\begin{eqnarray}
&&{}^{(1)}u^{a}=\left(\begin{array}{c} c_{10} \\ 0 \\ 0\\ 0\\ \end{array} \right),
{}^{(2)}u^{a}=\left(\begin{array}{c}  c_{20} \\ c_{21} \\ 0 \\ 0\\ \end{array} \right),
{}^{(3)}u^{a}=\left(\begin{array}{c}  c_{30} \\ 0 \\ c_{32} \\ 0 \\ \end{array} \right), \nonumber \\
&&{}^{(4)}u^{a}=\left(\begin{array}{c}  c_{40} \\ 0 \\ 0 \\ c_{43}\\ \end{array} \right),
{}^{(5)}u^{a}=\left(\begin{array}{c}  c_{50} \\ c_{51} \\ c_{52} \\ 0\\ \end{array} \right),
{}^{(6)}u^{a}=\left(\begin{array}{c}  c_{60} \\ 0 \\ c_{62} \\ c_{63} \\ \end{array} \right).\label{velocity_setup}
\end{eqnarray}
The $c_{(m)a}$ here are just constants, chosen appropriately to ensure the normalization of the 4-velocity of each compass. 

In summary, we are going to consider $(m)=1 \dots 6$ compasses, each of them with 6-points, where the five reference points are always the $(n)=1 \dots 5$ from (\ref{position_setup}).

\paragraph{Explicit curvature components}

The 20 independent components of the curvature tensor can be explicitly determined in terms of the accelerations ${}^{(m,n)}A_a$ and velocities ${}^{(m)}u^a$ by making use of the deviation equation (\ref{compass_lower_setup}) with the help of the compass configuration given in (\ref{position_setup}) and (\ref{velocity_setup}). The result reads as follows:
\begin{eqnarray}
01 : R_{1010} &=& \frac{3}{4} {}^{(1,1)}A_1 c^{-2}_{10}, \label{exR1010} \\
02 : R_{2010} &=& \frac{3}{4} {}^{(1,1)}A_2 c^{-2}_{10}, \label{exR2010} \\
03 : R_{3010} &=& \frac{3}{4} {}^{(1,1)}A_3 c^{-2}_{10}, \label{exR3010} \\
04 : R_{2020} &=& \frac{3}{4} {}^{(1,2)}A_2 c^{-2}_{10}, \label{exR2020} \\
05 : R_{3020} &=& \frac{3}{4} {}^{(1,2)}A_3 c^{-2}_{10}, \label{exR3020} \\
06 : R_{3030} &=& \frac{3}{4} {}^{(1,3)}A_3 c^{-2}_{10}, \label{exR3030} \\
07 : R_{2110} &=& \frac{3}{4} {}^{(2,1)}A_2 c^{-1}_{21} c^{-1}_{20} -  R_{2010} c^{-1}_{21} c_{20} , \label{exR2110} \\
08 : R_{3110} &=& \frac{3}{4} {}^{(2,1)}A_3 c^{-1}_{21} c^{-1}_{20} -  R_{3010} c^{-1}_{21} c_{20} , \label{exR3110} \\
09 : R_{0212} &=& \frac{3}{4} {}^{(3,1)}A_0 c^{-2}_{32} +  R_{2010} c^{-1}_{32} c_{30} , \label{exR0212} \\
10 : R_{1212} &=& \frac{3}{4} {}^{(2,2)}A_2 c^{-2}_{21} -  R_{2020} c^{2}_{20} c^{-2}_{21}  -  2 R_{0212} c^{-1}_{21} c_{20}  , \label{exR1212} \\
11 : R_{3220} &=& \frac{3}{4} {}^{(3,2)}A_3 c^{-1}_{32} c^{-1}_{30} -  R_{3020} c^{-1}_{32} c_{30} , \label{exR3220} \\
12 : R_{0313} &=& \frac{3}{4} {}^{(4,1)}A_0 c^{-2}_{43} +  R_{3010} c^{-1}_{43} c_{40} , \label{exR0313} \\
13 : R_{1313} &=& \frac{3}{4} {}^{(2,3)}A_3 c^{-2}_{21} -  R_{3030} c^{2}_{20} c^{-2}_{21}  -  2 R_{0313} c^{-1}_{21} c_{20}  , \label{exR1313} \\
14 : R_{0323} &=& \frac{3}{4} {}^{(4,2)}A_0 c^{-2}_{43} +  R_{3020} c^{-1}_{43} c_{40} , \label{exR0323}\\
15 : R_{2323} &=& \frac{3}{4} {}^{(4,2)}A_2 c^{-2}_{43} -  R_{2020} c^{-2}_{43} c^{2}_{40}  +  2 R_{3220} c^{-1}_{43} c_{40}  , \label{exR2323} 
\end{eqnarray}
\begin{eqnarray}
16 : R_{3132} &=& \frac{3}{8} {}^{(5,3)}A_3 c^{-1}_{52} c^{-1}_{51}  -  \frac{1}{2} R_{3030} c^{-1}_{52} c^{-1}_{51} c^2_{50} \nonumber \\ 
&&  -  R_{0313} c^{-1}_{52} c_{50} -  R_{0323} c^{-1}_{51} c_{50} -  \frac{1}{2} R_{1313} c^{-1}_{52} c_{51} \nonumber \\
&&  - \frac{1}{2} R_{2323} c_{52} c^{-1}_{51} ,  \label{exR3132} \\
17 : R_{1213} &=& \frac{3}{8} {}^{(6,1)}A_1 c^{-1}_{63} c^{-1}_{62}  -  \frac{1}{2} R_{1010} c^{-1}_{63} c^{-1}_{62} c^2_{60} \nonumber \\ 
&&  +  R_{2110} c^{-1}_{63} c_{60} +  R_{3110} c^{-1}_{62} c_{60} -  \frac{1}{2} R_{1212} c^{-1}_{63} c_{62} \nonumber \\
&&  - \frac{1}{2} R_{1313} c_{63} c^{-1}_{62} ,  \label{exR1213} 
\end{eqnarray}

\begin{figure}
  \begin{minipage}[c]{0.55\textwidth}
    \includegraphics[width=\textwidth,angle=-90]{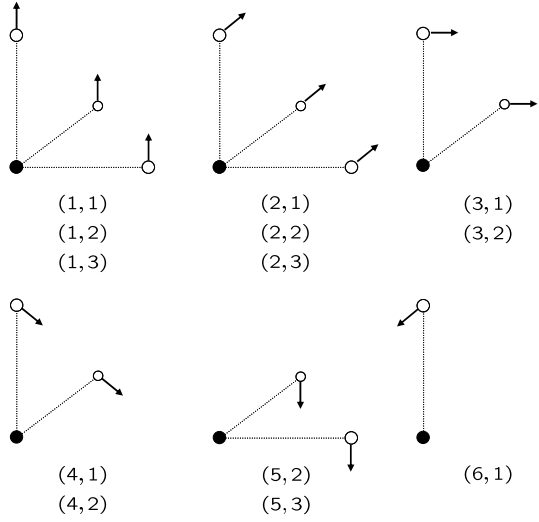}
  \end{minipage}\hfill
  \begin{minipage}[c]{0.4\textwidth}
   \caption{\label{fig_standard_compass} Symbolical sketch of the explicit compass solution in (\ref{exR1010})-(\ref{ex3212}). In total 13 suitably prepared test bodies (hollow circles) are needed to determine all curvature components. The reference body is denoted by the black circle. Note that with the standard deviation equation all ${}^{(1 \dots 6)}u^{a}$, but only ${}^{(1 \dots 3)}\eta^{a}$ are needed in the solution.}
  \end{minipage}
\end{figure}

There are still 3 components of the curvature tensor missing. To determine them, we notice that the following relation between the remaining equations is at our disposal:
\begin{eqnarray}
R_{0312}-R_{0231} &=& \frac{3}{4} {}^{(2,2)}A_3 c^{-1}_{20} c^{-1}_{21} - R_{3020} c_{20} c^{-1}_{21}
- R_{3121} c_{21} c^{-1}_{20}, \label{difference_1} \\
R_{0231}-R_{0123} &=& \frac{3}{4} {}^{(4,1)}A_2 c^{-1}_{40} c^{-1}_{43} - R_{2010} c_{40} c^{-1}_{43}
- R_{2313} c_{43} c^{-1}_{40}. \label{difference_2}
\end{eqnarray}
Subtracting (\ref{difference_1}) from (\ref{difference_2}) and using the Ricci identity we find:
\begin{eqnarray}
18 : R_{0231} &=& \frac{1}{4} {}^{(4,1)}A_2 c^{-1}_{40} c^{-1}_{43}  -  \frac{1}{4} {}^{(2,2)}A_3 c^{-1}_{20} c^{-1}_{21} \nonumber \\ 
&&  + \frac{1}{3} \big( R_{3020} c_{20} c^{-1}_{21} + R_{3121} c_{21} c^{-1}_{20} \nonumber \\
&&  -  R_{2010} c_{40} c^{-1}_{43} -  R_{2313} c_{43} c^{-1}_{40} \big),  \label{ex0231} \\ 
19 : R_{0312} &=& \frac{1}{4} {}^{(4,1)}A_2 c^{-1}_{40} c^{-1}_{42}  + \frac{1}{2} {}^{(2,2)}A_3 c^{-1}_{20} c^{-1}_{21} \nonumber \\ 
&&  - \frac{1}{3} \big( 2 R_{3020} c_{20} c^{-1}_{21} + 2 R_{3121} c_{21} c^{-1}_{20} \nonumber \\
&&  + R_{2010} c_{40} c^{-1}_{43} +  R_{2313} c_{43} c^{-1}_{40}  \big).  \label{ex0312}
\end{eqnarray}
Finally, by reinsertion of (\ref{difference_1}) in one of the remaining compass equations, one obtains:
\begin{eqnarray}
20 : R_{3212} &=& \frac{3}{4} {}^{(4,1)}A_3  c^{-1}_{20}  c^{-1}_{21}  c_{50}  c^{-1}_{52}  -  \frac{3}{4} {}^{(5,2)}A_3  c^{-1}_{51}  c^{-1}_{52}  \nonumber \\ 
&& + R_{3121}  c^{-1}_{52} \left(c_{51} - c_{50}  c_{21} c^{-1}_{20}\right)  + R_{3220} c_{50} c^{-1}_{51} \nonumber \\
&&  + R_{3020}  c_{50}  c^{-1}_{52} \left(c_{50} c^{-1}_{51} -c_{20}  c^{-1}_{21} \right) .  \label{ex3212}
\end{eqnarray}
By examination of the components given in (\ref{exR1010})-(\ref{ex3212}), we conclude that for a full determination of the curvature one needs 13 test bodies, see fig.\ \ref{fig_standard_compass} for a sketch of the solution.

%
%\begin{figure}
%\begin{center}
%\includegraphics[width=6.4cm,angle=-90]{deviation_fig2.eps}
%\end{center}
%\caption{\label{fig_standard_compass} Symbolical sketch of the explicit compass solution in (\ref{exR1010})-(\ref{ex3212}). In total 13 suitably prepared test bodies (hollow circles) are needed to determine all curvature components. The reference body is denoted by the black circle. Note that with the standard deviation equation all ${}^{(1 \dots 6)}u^{a}$, but only ${}^{(1 \dots 3)}\eta^{a}$ are needed in the solution.}
%\end{figure}

\paragraph{Vacuum spacetime}

\begin{figure}
  \begin{minipage}[c]{0.55\textwidth}
    \includegraphics[width=\textwidth,angle=-90]{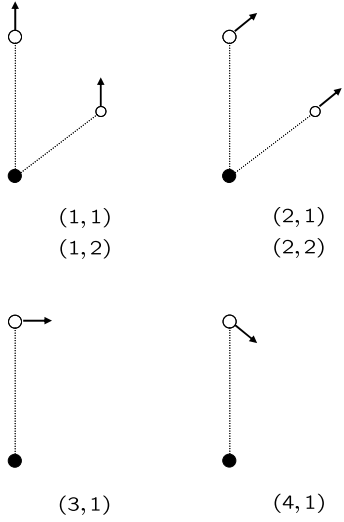}
  \end{minipage}\hfill
  \begin{minipage}[c]{0.4\textwidth}
   \caption{\label{fig_standard_compass_vacuum} Symbolical sketch of the explicit compass solution in (\ref{exC1010})-(\ref{exC0312}) for the vacuum case. In total 6 suitably prepared test bodies (hollow circles) are needed to determine all components of the Weyl tensor. The reference body is denoted by the black circle. Note that in vacuum, with the standard deviation equation, all ${}^{(1 \dots 4)}u^{a}$, but only ${}^{(1 \dots 2)}\eta^{a}$ are needed in the solution.}
  \end{minipage}
\end{figure}

In vacuum the number of independent components of the curvature is reduced to the 10 components of the Weyl tensor $C_{abcd}$. Replacing $R_{abcd}$ in the compass solution (\ref{exR1010})-(\ref{ex3212}), and taking into account the symmetries of Weyl we may use a reduced compass setup to completely determine the gravitational field, i.e.\  

\begin{eqnarray}
01 : C_{1010} &=& \frac{3}{4} {}^{(1,1)}A_1 c^{-2}_{10}, \label{exC1010} \\
02 : C_{2010} &=& \frac{3}{4} {}^{(1,1)}A_2 c^{-2}_{10}, \label{exC2010} \\
03 : C_{3010} &=& \frac{3}{4} {}^{(1,1)}A_3 c^{-2}_{10}, \label{exC3010} \\
04 : C_{2020} &=& \frac{3}{4} {}^{(1,2)}A_2 c^{-2}_{10}, \label{exC2020} \\
05 : C_{3020} &=& \frac{3}{4} {}^{(1,2)}A_3 c^{-2}_{10}, \label{exC3020} \\
06 : C_{2110} &=& \frac{3}{4} {}^{(2,1)}A_2 c^{-1}_{21} c^{-1}_{20} -  C_{2010} c^{-1}_{21} c_{20} , \label{exC2110} \\
07 : C_{3110} &=& \frac{3}{4} {}^{(2,1)}A_3 c^{-1}_{21} c^{-1}_{20} -  C_{3010} c^{-1}_{21} c_{20} , \label{exC3110} \\
08 : C_{0212} &=& \frac{3}{4} {}^{(3,1)}A_0 c^{-2}_{32} +  C_{2010} c^{-1}_{32} c_{30} , \label{exC0212}
\end{eqnarray}
\begin{eqnarray}
09 : C_{0231} &=& \frac{1}{4} {}^{(4,1)}A_2 c^{-1}_{40} c^{-1}_{43}  -  \frac{1}{4} {}^{(2,2)}A_3 c^{-1}_{20} c^{-1}_{21} \nonumber \\ 
&&  + \frac{1}{3} C_{3020} \big(  c_{20} c^{-1}_{21} + c_{21} c^{-1}_{20}\big) \nonumber \\
&&  - \frac{1}{3} C_{2010}\big( c_{40} c^{-1}_{43} + c_{43} c^{-1}_{40} \big),  \label{exC0231} \\ 
10 : C_{0312} &=& \frac{1}{4} {}^{(4,1)}A_2 c^{-1}_{40} c^{-1}_{42}  + \frac{1}{2} {}^{(2,2)}A_3 c^{-1}_{20} c^{-1}_{21} \nonumber \\ 
&&  - \frac{2}{3} C_{3020} \big( c_{20} c^{-1}_{21} + c_{21} c^{-1}_{20} \big) \nonumber \\
&&  + \frac{1}{3} C_{2010} \big( c_{40} c^{-1}_{43} + c_{43} c^{-1}_{40} \big). \label{exC0312}
\end{eqnarray}
All the other components of the Weyl tensor are obtained from the above by making use of the double-self-duality property $C_{abcd} = -\frac{1}{4}\epsilon_{abef}\epsilon_{cdgh}C^{efgh}$, where $\epsilon_{abcd}$ is the totally antisymmetric Levi-Civita tensor with $\epsilon_{0123}=1$, and the Ricci identity. See fig.\ \ref{fig_standard_compass_vacuum} for a sketch of the  solution.

%
%\begin{figure}
%\begin{center}
%\includegraphics[width=7cm,angle=-90]{deviation_fig3.eps}
%\end{center}
%\caption{\label{fig_standard_compass_vacuum} Symbolical sketch of the explicit compass solution in (\ref{exC1010})-(\ref{exC0312}) for the vacuum case. In total 6 suitably prepared test bodies (hollow circles) are needed to determine all components of the Weyl tensor. The reference body is denoted by the black circle. Note that in vacuum, with the standard deviation equation, all ${}^{(1 \dots 4)}u^{a}$, but only ${}^{(1 \dots 2)}\eta^{a}$ are needed in the solution.}
%\end{figure}

\subsection{Method 2: Relativistic clock gradiometry / Gravitational clock compass }\label{paragraph_fermi_normal_curvature_determination}

Now we turn to the determination of the curvature in a general spacetime by means of clocks. We consider the non-vacuum case first, when one needs to measure 20 independent components of the Riemann curvature tensor $R_{abc}{}^d$. 

\begin{figure}
\begin{center}
\includegraphics[width=5.5cm,angle=-90]{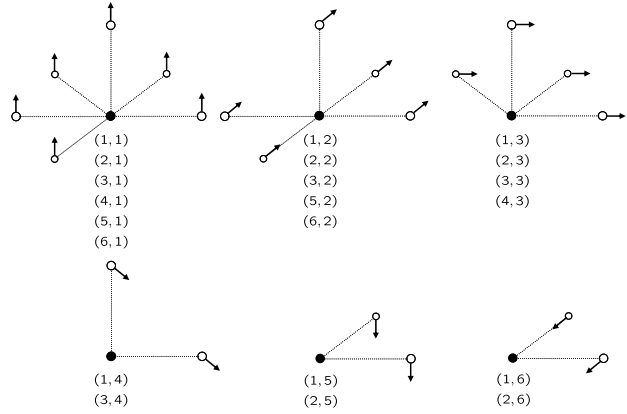}
\end{center}
\caption{\label{fig_5} Symbolical sketch of the explicit solution for the curvature (\ref{curvature_1})-(\ref{curvature_21}). In total 21 suitably prepared clocks (hollow circles) are needed to determine all curvature components. The observer is denoted by the black circle. Note that all ${}^{(1 \dots 6)}v^{a}$, but only ${}^{(1 \dots 3)}y^{a}$ are needed in the solution.}
\end{figure}

Again we start by rearranging the system (\ref{general_freq_ratio_definition}):
\begin{eqnarray}
&& {}^{(n)}y^\alpha  {}^{(n)}y^\beta \left(-R_{0 \alpha \beta 0} - \frac{4}{3} R_{\gamma \alpha \beta 0} \, {}^{(m)}v^\gamma + \frac{1}{3} R_{ \alpha \gamma \delta \beta } \, {}^{(m)}v^\gamma {}^{(m)}v^\delta \right) \nonumber \\
&&= B({}^{(n)}y^\alpha, {}^{(m)}v^\alpha, {}^{(p)}a^\alpha, {}^{(q)}\omega^\alpha), \label{curvature_master}
\end{eqnarray}
where
\begin{eqnarray}
B(y^\alpha, v^\alpha, a^\alpha, \omega^\alpha) &:=& (1 - v^2) \left( C-1 \right) - 2 a_\alpha y^\alpha - y^\alpha y^\beta \Bigg( a_\alpha a_\beta  \nonumber \\
&&- \delta_{\alpha \beta} \omega_\gamma \omega^\gamma + \omega_\alpha \omega_\beta \Bigg) - 2 v^\alpha  \varepsilon_{\alpha \beta \gamma}  y^\beta \omega^\gamma . \label{rhs_curvature_determination} 
\end{eqnarray}
Analogously to our analysis of the gravitational compass \cite{Puetzfeld:Obukhov:2016:1}, we may now consider different setups of clocks to measure as many curvature components as possible. The system in (\ref{curvature_master}) yields (please note that only the position and the velocity indices are indicated):
\begin{eqnarray}
01 : R_{1010}&=& {}^{(1,1)}B, \label{curvature_1} \\
02 : R_{2110}&=& \frac{3}{4}{c^{-1}_{22}c^{-1}_{42}\left(c_{22}-c_{42}\right)^{-1}} \Bigg({}^{(1,1)}B c_{22}^2 - {}^{(1,1)}B c_{42}^2 \nonumber \\
&&+ {}^{(1,2)}B c_{42}^2 - {}^{(1,4)}B c_{22}^2  \Bigg) , \label{curvature_2} \\
03 : R_{1212}&=& -3 {c^{-1}_{22}c^{-1}_{42}\left(c_{22}-c_{42}\right)^{-1}} \Bigg({}^{(1,1)}B c_{22} - {}^{(1,1)}B c_{42} \nonumber \\
&&+ {}^{(1,2)}B c_{42} - {}^{(1,4)}B c_{22}  \Bigg), \label{curvature_3} \\
04 : R_{3110}&=& \frac{3}{4}{c^{-1}_{33}c^{-1}_{63}\left(c_{33}-c_{63}\right)^{-1}} \Bigg({}^{(1,1)}B c_{33}^2 - {}^{(1,1)}B c_{63}^2 \nonumber \\
&& + {}^{(1,3)}B c_{63}^2 - {}^{(1,6)}B c_{33}^2  \Bigg), \label{curvature_4} \\
05 : R_{1313}&=& -3 {c^{-1}_{33}c^{-1}_{63}\left(c_{33}-c_{63}\right)^{-1}}\Bigg({}^{(1,1)}B c_{33} - {}^{(1,1)}B c_{63} \nonumber \\
&&+ {}^{(1,3)}B c_{63} - {}^{(1,6)}B c_{33}  \Bigg), \label{curvature_5}
\end{eqnarray}
\begin{eqnarray}
06 : R_{1213}&=& \frac{3}{2} c^{-1}_{52} c^{-1}_{53} \Bigg(-\,{}^{(1,5)}B + R_{1010} - \frac{4}{3}R_{2110}c_{52} \nonumber \\
&& - \frac{4}{3}R_{3110} c_{53} - \frac{1}{3}R_{1212} c^2_{52}- \frac{1}{3}R_{1313} c^2_{53} \Bigg), \label{curvature_6}\\
07 : R_{2020}&=& {}^{(2,2)}B, \label{curvature_7}\\
08 : R_{0212}&=& \frac{3}{4} c^{-1}_{11} \Bigg({}^{(2,1)}B - R_{2020} +\frac{1}{3} R_{1212}c_{11}^2 \Bigg), \label{curvature_8}\\
09 : R_{3220}&=& \frac{3}{4}{c^{-1}_{33}c^{-1}_{53}\left(c_{33}-c_{53}\right)^{-1}} \Bigg({}^{(2,2)}B c_{33}^2 - {}^{(2,2)}B c_{53}^2 \nonumber \\
&&+ {}^{(2,3)}B c_{53}^2 - {}^{(2,5)} B c_{33}^2  \Bigg), \label{curvature_9}\\
10 : R_{2323}&=& -3{c^{-1}_{33}c^{-1}_{53}\left(c_{33}-c_{53}\right)^{-1}}\Bigg({}^{(2,2)}B c_{33} - {}^{(2,2)}B c_{53} \nonumber \\
&&+ {}^{(2,3)}B c_{53} - {}^{(5,2)}B c_{33}  \Bigg), \label{curvature_10}\\
11 : R_{3212}&=& \frac{3}{2} c^{-1}_{61} c^{-1}_{63} \Bigg( -\,{}^{(2,6)}B + R_{2020} + \frac{4}{3}R_{0212}c_{61}- \frac{4}{3} R_{3220} c_{63} \nonumber \\
&&- \frac{1}{3} R_{1212} c^2_{61}-\frac{1}{3}R_{2323}c^2_{63}\Bigg), \label{curvature_11} \\
12 : R_{3030}&=& {}^{(3,3)}B, \label{curvature_12}\\
13 : R_{0313}&=& \frac{3}{4} c^{-1}_{11} \Bigg({}^{(3,1)}B - R_{3030} +\frac{1}{3} R_{1313}c_{11}^2 \Bigg), \label{curvature_13}\\
14 : R_{0323}&=& \frac{3}{4}c^{-1}_{22} \Bigg({}^{(3,2)}B - R_{3030} +\frac{1}{3} R_{2323}c_{22}^2 \Bigg), \label{curvature_14} \\ 
15 : R_{3132}&=& \frac{3}{2} c^{-1}_{41} c^{-1}_{42} \Bigg(-\,{}^{(3,4)}B + R_{3030} + \frac{4}{3}R_{0313} c_{41} + \frac{4}{3} R_{0323} c_{42} \nonumber \\
&&- \frac{1}{3} R_{1313} c^2_{41} - \frac{1}{3}R_{2323}c^{2}_{42}\Bigg), \label{curvature_15}
\end{eqnarray}
\begin{eqnarray}
16 : R_{2010}&=& \frac{1}{2} \Bigg({}^{(4,1)}B - R_{1010} - R_{2020} - \frac{4}{3} R_{0212}c_{11} \nonumber \\
&&- \frac{4}{3} R_{2110}c_{11} +\frac{1}{3} R_{1212}c^{2}_{11} \Bigg), \label{curvature_16} \\
17 : R_{3020}&=& \frac{1}{2} \Bigg({}^{(5,2)}B - R_{2020} - R_{3030} - \frac{4}{3}R_{0323}c_{22} \nonumber \\
&&- \frac{4}{3}R_{3220}c_{22}+\frac{1}{3}R_{2323}c^{2}_{22}\Bigg), \label{curvature_17} \\
18 : R_{3010}&=& \frac{1}{2} \Bigg({}^{(6,1)}B - R_{1010} - R_{3030} - \frac{4}{3} R_{0313}c_{11}\nonumber \\
&& - \frac{4}{3} R_{3110}c_{11}+\frac{1}{3} R_{1313}c^{2}_{11}\Bigg). \label{curvature_18}
\end{eqnarray}
Introducing abbreviations 
\begin{eqnarray}
K_1&:=& \frac{3}{4} c_{33}^{-1}\Bigg[-\,{}^{(4,3)}B+R_{1010} + 2R_{2010} + R_{2020} \nonumber \\
&&- \frac{4}{3}(R_{3110} + R_{3220})c_{33}-\frac{1}{3}(R_{1313} + 2R_{3132} + R_{2323})c^2_{33} \Bigg],\label{k1_definition} \\
K_2&:=& \frac{3}{4}c_{11}^{-1}\Bigg[-\,{}^{(5,1)}B + R_{2020} + 2R_{3020} + R_{3030} \nonumber \\
&&+\frac{4}{3}(R_{0212} + R_{0313})c_{11}-\frac{1}{3}(R_{1212} + 2R_{1213} + R_{1313})c^2_{11} \Bigg],\label{k2_definition} \\
K_3&:=&\frac{3}{4} c_{22}^{-1}\Bigg[-\,{}^{(6,2)}B + R_{1010} + 2R_{3010} + R_{3030} \nonumber \\
&&-\frac{4}{3}(R_{2110} + R_{0323})c_{22} - \frac{1}{3}(R_{1212} + 2R_{3212}+ R_{2323})c^2_{22} \Bigg],\label{k3_definition}
\end{eqnarray}
we find the remaining three curvature components 
\begin{eqnarray}
19 : R_{1023}&=& \frac{1}{3} \left(K_3 - K_1\right) , \label{curvature_19}
\end{eqnarray}
\begin{eqnarray}
20 : R_{2013}&=& \frac{1}{3} \left(K_2 - K_1 \right) , \label{curvature_20}\\
21 : R_{3021}&=& \frac{1}{3} \left(K_3 - K_2\right) . \label{curvature_21}
\end{eqnarray}
See figure \ref{fig_5} for a symbolical sketch of the solution. The $B$'s in these equations can be explicitly resolved in terms of the $C$'s
\begin{eqnarray}
{}^{(1,1)}B&=&\left( 1 - c_{11}^2 \right) \left( {}^{(1,1)}C-1 \right), \label{B11_curved_case} \\
{}^{(1,2)}B&=&\left( 1 - c_{22}^2 \right) \left( {}^{(1,2)}C-1 \right), \label{B12_curved_case} \\
{}^{(1,3)}B&=&\left( 1 - c_{33}^2 \right) \left( {}^{(1,3)}C-1 \right), \label{B13_curved_case} \\
{}^{(1,4)}B&=&\left( 1 - c_{41}^2 - c_{42}^2 \right) \left( {}^{(1,4)}C-1 \right), \label{B14_curved_case}\\
{}^{(1,5)}B&=&\left( 1 - c_{52}^2 - c_{53}^2 \right) \left( {}^{(1,5)}C-1 \right), \label{B15_curved_case} \\
{}^{(1,6)}B&=&\left( 1 - c_{61}^2 - c_{63}^2 \right) \left( {}^{(1,6)}C-1 \right), \label{B16_curved_case}\\
{}^{(2,1)}B&=&\left( 1 - c_{11}^2 \right) \left( {}^{(2,1)}C-1 \right), \label{B21_curved_case} \\
{}^{(2,2)}B&=&\left( 1 - c_{22}^2 \right) \left( {}^{(2,2)}C-1 \right), \label{B22_curved_case} \\
{}^{(2,3)}B&=&\left( 1 - c_{33}^2 \right) \left( {}^{(2,3)}C-1 \right), \label{B23_curved_case} \\
{}^{(2,5)}B&=&\left( 1 - c_{52}^2 - c_{53}^2 \right) \left( {}^{(2,5)}C-1 \right), \label{B25_curved_case} \\
{}^{(2,6)}B&=&\left( 1 - c_{61}^2 - c_{63}^2 \right) \left( {}^{(2,6)}C-1 \right), \label{B26_curved_case} \\
{}^{(3,1)}B&=&\left( 1 - c_{11}^2 \right) \left( {}^{(3,1)}C-1 \right), \label{B31_curved_case}\\
{}^{(3,2)}B&=&\left( 1 - c_{22}^2 \right) \left( {}^{(3,2)}C-1 \right), \label{B32_curved_case} \\
{}^{(3,3)}B&=&\left( 1 - c_{33}^2 \right) \left( {}^{(3,3)}C-1 \right), \label{B33_curved_case} \\
{}^{(3,4)}B&=&\left( 1 - c_{41}^2 - c_{42}^2 \right) \left( {}^{(3,4)}C-1 \right), \label{B34_curved_case}\\
{}^{(4,1)}B&=&\left( 1 - c_{11}^2 \right) \left( {}^{(4,1)}C-1 \right), \label{B41_curved_case}\\
{}^{(4,3)}B&=&\left( 1 - c_{33}^2 \right) \left( {}^{(4,1)}C-1 \right), \label{B43_curved_case} \\
{}^{(5,1)}B&=&\left( 1 - c_{11}^2 \right) \left( {}^{(5,1)}C-1 \right), \label{B51_curved_case} \\
{}^{(5,2)}B&=&\left( 1 - c_{22}^2 \right) \left( {}^{(5,2)}C-1 \right), \label{B52_curved_case}
\end{eqnarray}
\begin{eqnarray} 
{}^{(6,1)}B&=&\left( 1 - c_{11}^2 \right) \left( {}^{(6,1)}C-1 \right), \label{B61_curved_case} \\
{}^{(6,2)}B&=&\left( 1 - c_{22}^2 \right) \left( {}^{(6,2)}C-1 \right). \label{B62_curved_case} 
\end{eqnarray}

\paragraph{Vacuum spacetime}\label{subsection_vacuum_spacetime}

\begin{figure}
\begin{center}
\includegraphics[width=5.5cm,angle=-90]{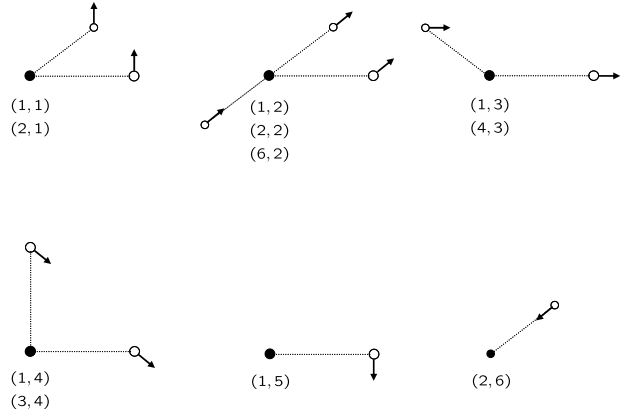}
\end{center}
\caption{\label{fig_6} Symbolical sketch of the explicit vacuum solution for the curvature (\ref{vacuum_curvature_1})-(\ref{vacuum_curvature_12}). In total 11 suitably prepared clocks (hollow circles) are needed to determine all curvature components. The observer is denoted by the black circle. Note that all ${}^{(1 \dots 6)}v^{a}$, but only ${}^{(1 \dots 3)}y^{a}$ are needed in the solution.}
\end{figure}

In vacuum the number of independent components of the curvature is reduced to the 10 components of the Weyl tensor $C_{abcd}$. Replacing $R_{abcd}$ in the compass solution (\ref{curvature_1})-(\ref{curvature_18}), and taking into account the symmetries of the Weyl tensor, we may use a reduced clock setup to completely determine the gravitational field. All other components may be obtained from the double self-duality property $C_{abcd}=-\frac{1}{4}\varepsilon_{abef}\varepsilon_{cdgh}C^{efgh}$.
\begin{eqnarray}
01 : C_{2323}&=& - {}^{(1,1)}B, \label{vacuum_curvature_1} \\
02 : C_{0323}&=& \frac{3}{4}{c^{-1}_{22}c^{-1}_{42}\left(c_{22}-c_{42}\right)^{-1}} \Bigg({}^{(1,1)}B c_{22}^2 - {}^{(1,1)}B c_{42}^2 \nonumber \\
&&+ {}^{(1,2)}B c_{42}^2 - {}^{(1,4)}B c_{22}^2  \Bigg) , \label{vacuum_curvature_2}
\end{eqnarray}
\begin{eqnarray}
03 : C_{3030}&=& 3 {c^{-1}_{22}c^{-1}_{42}\left(c_{22}-c_{42}\right)^{-1}} \Bigg({}^{(1,1)}B c_{22} - {}^{(1,1)}B c_{42}  \nonumber \\
&&+ {}^{(1,2)}B c_{42} - {}^{(1,4)}B c_{22}  \Bigg), \label{vacuum_curvature_3}\\
04 : C_{2020}&=& {}^{(2,2)}B, \label{vacuum_curvature_4} \\
05 : C_{3220}&=& \frac{3}{4}c^{-1}_{33} \left({}^{(1,3)}B + C_{2323} - \frac{1}{3} C_{2020}c_{33}^2 \right), \label{vacuum_curvature_5} \\
06 : C_{0313}&=& - \frac{3}{4}c^{-1}_{11} \left({}^{(2,1)}B -  C_{2020} - \frac{1}{3} C_{3030}c_{11}^2 \right), \label{vacuum_curvature_6} \\
07 : C_{3020}&=& - \frac{3}{2}c^{-1}_{52}c^{-1}_{53} \Bigg({}^{(1,5)}B +  C_{2323} + \frac{4}{3} C_{0323}c_{52}  \nonumber \\
&&- \frac{4}{3} C_{3220}c_{53} - \frac{1}{3} C_{3030}c^2_{52} - \frac{1}{3} C_{2020}c^2_{53}  \Bigg), \label{vacuum_curvature_7} \\
08 : C_{3212}&=& - \frac{3}{2}c^{-1}_{61}c^{-1}_{63} \Bigg({}^{(2,6)}B -  C_{2020} + \frac{4}{3} C_{0313}c_{61} + \frac{4}{3} C_{3220}c_{63}  \nonumber \\
&&- \frac{1}{3} C_{3030}c^2_{61} + \frac{1}{3} C_{2323}c^2_{63}  \Bigg), \label{vacuum_curvature_8} \\
09 : C_{3132}&=& - \frac{3}{2}c^{-1}_{41}c^{-1}_{42} \Bigg({}^{(3,4)}B -  C_{3030} - \frac{4}{3} C_{0313}c_{41} - \frac{4}{3} C_{0323}c_{42}  \nonumber \\
&&- \frac{1}{3} C_{2020}c^2_{41} + \frac{1}{3} C_{2323}c^2_{42}  \Bigg). \label{vacuum_curvature_9}
\end{eqnarray}
With the abbreviations 
\begin{eqnarray}
K_1&:=& \frac{3}{4} c_{33}^{-1}\Bigg[-{}^{(4,3)}B-C_{2323} + 2C_{3132} + C_{2020}  \nonumber \\
&&+ \frac{1}{3}(C_{2020} - 2C_{3132} - C_{2323})c^2_{33} \Bigg],\label{k1_definition_2} 
\end{eqnarray}
\begin{eqnarray}
K_2&:=& \frac{3}{4}c_{11}^{-1}\Bigg[-{}^{(5,1)}B + C_{2020} + 2C_{3020} + C_{3030}  \nonumber \\
&&+ \frac{1}{3}(C_{3030} - 2C_{3020} + C_{2020})c^2_{11} \Bigg],\label{k2_definition_2} \\
K_3&:=&-\frac{3}{4} c_{22}^{-1}\Bigg[{}^{(6,2)}B + C_{2323} - 2C_{3212} - C_{3030}   \nonumber \\
&&- \frac{1}{3}(C_{3030} - 2C_{3212}- C_{2323})c^2_{22} \Bigg],\label{k3_definition_2}
\end{eqnarray}
the remaining three curvature components read
\begin{eqnarray}
10 : C_{1023}&=& \frac{1}{3} \left(K_3 - K_1\right), \label{vacuum_curvature_10}\\
11 : C_{2013}&=& \frac{1}{3} \left(K_2 - K_1\right), \label{vacuum_curvature_11}\\
12 : C_{3021}&=& \frac{1}{3} \left(K_3 - K_2\right). \label{vacuum_curvature_12}
\end{eqnarray}
A symbolical sketch of the solution is given in figure \ref{fig_6}.

\section{Summary}\label{sec_summary}

\subsection{Method 1: Summary}\label{sec_method_1_summary}

In the framework of Synge's world function approach, we have derived a generalized covariant deviation equation (\ref{eta_2nd_deriv_alternative_2}) which is valid for arbitrary world lines and in general background spacetimes. Making use of systematic expansions of the exact deviation equation up to the third order in the world function, we obtain the final result (\ref{eta_2nd_deriv_compact_2}) which can be viewed as a generalization of the well-known geodesic deviation equation. Furthermore, our results encompass several suggestions for a generalized deviation equation from the literature as special cases, and therefore may serve a unified framework for further studies.  

In section \ref{sec_compass} we have shown how deviation equations can be used to determine the curvature of spacetime. For this we extended the notion of a gravitational compass \cite{Szekeres:1965} and worked out compass setups for general as well as for vacuum spacetimes. One setup is based on the standard geodesic deviation equation, and another is based on the next order generalization given in which goes beyond the linearized case. For both cases we provided the explicit compass solution which allows for a full determination of the curvature. 

In contrast to the general considerations in \cite{Synge:1960,Szekeres:1965} we give an explicit exact solution for the compass setup. With the standard deviation equation, as well as with the generalized deviation equation, we need at least 13 test bodies to determine all curvature components in a general spacetime. For the standard deviation we therefore obtain the same number of bodies as in \cite{Ciufolini:Demianski:1986}, however it is worthwhile to note that no explicit solution was given in \cite{Ciufolini:Demianski:1986} for a non-vacuum spacetime. In the case of a generalized deviation equation our findings are at odds with the results in \cite{Ciufolini:Demianski:1986}. However, this discrepancy in the generalized case comes as no surprise since the generalized equation used in \cite{Ciufolini:Demianski:1986} -- which was previously derived in \cite{Ciufolini:1986} -- differs from our equation. In vacuum spacetimes, we have explicitly shown that the number of required test bodies is reduced to 6, for the standard deviation equation, and to 5, for the generalized deviation equation. 

Furthermore, it is interesting to note that in the case of the standard deviation equation, the opinion of the authors \cite{Synge:1960,Szekeres:1965} differs when it comes to the number of required test bodies. This seems to be related to the counting scheme and the interpretation of the notion of a compass. Since no explicit compass solutions were given in \cite{Synge:1960,Szekeres:1965}, one cannot make a comparison to our results. In the case of \cite{Ciufolini:Demianski:1986}, we were not able to verify that the given solution does fulfill the compass equations derived in that work. However, the agreement on the number of required bodies in combination with the standard deviation is reassuring. 

Our results are of direct operational relevance and form the basis for many experiments. Important applications range from the description of gravitational wave detectors to the study of satellite configurations for gravitational field mapping in relativistic geodesy. 

\subsection{Method 2: Summary}\label{sec_method_2_summary}

Section \ref{paragraph_fermi_normal_curvature_determination} describes an experimental setup which we call a clock compass, in analogy to the usual gravitational compass \cite{Szekeres:1965,Puetzfeld:Obukhov:2016:1}. We have shown that a suitably prepared set of clocks can be used to determine all components of the gravitational field, i.e.\ the curvature, in General Relativity, as well as to describe the state of motion of a noninertial observer.  

Working out explicit clock compass setups in different situations, we have demonstrated that in general 6 clocks are needed to determine the linear acceleration as well as the rotational velocity, while 4 clocks will suffice in case of the velocity. Furthermore, we prove that one needs 21 and 11 clocks, respectively, to determine all curvature components in a general curved spacetime and in vacuum. In view of possible future experimental realizations it is interesting to note that restrictions regarding the choice of clock velocities in a setup lead to restrictions regarding the number of determinable curvature components. 

Our results are of direct operational relevance for the setup of networks of clocks, especially in the context of relativistic geodesy. In geodetic terms, the given clock configurations may be thought of as a clock gradiometers. Taking into account the steadily increasing accuracy of clocks \cite{Rodrigo:etal:2018}, these results should be combined with those from a gradiometric context, for example in the form of a hybrid gravitational compass -- which combines acceleration as well as clock measurements in one setup. Another possible application is the detection of gravitational waves by means of clock as well as standard interferometric techniques. An interesting question is a possible reduction of the number of measurements by a combination of different techniques.

\section{Outlook: Operational determination of the gravitational field in theories beyond GR}\label{sec_outlook}

In the previous sections, we have shown how the deviation equation as well as an ensemble of clocks can be used to measure the gravitational field in GR. However, the results were limited to theories in a Riemannian background. While such theories are justified in many physical situations, several modern gravitational theories \cite{Hehl:1976,Blagojevic:Hehl:2013,Ponomarev:Bravinsky:Obukhov:2017} reach significantly beyond the Riemannian geometrical framework. In particular it is already well-known \cite{Hehl:Obukhov:Puetzfeld:2013,Puetzfeld:Obukhov:2014:2,Obukhov:Puetzfeld:2015:1}, that in the description of test bodies with intrinsic degrees of freedom -- like spin -- there is a natural coupling to the post-Riemannian features of spacetime. Therefore, in view of possible tests of gravitational theories by means of structured test bodies, a further extension of the deviation equation to post-Riemannian geometries is needed.

In the following we present a generalized deviation equation in a Riemann-Cartan background, allowing for spacetimes endowed with torsion, the presentation is based on \cite{Puetzfeld:Obukhov:2018:2}. This equation describes the dynamics of the connecting vector which links events on two general (adjacent) world lines. Our results are valid for any theory in a Riemann-Cartan background, in particular they apply to Einstein-Cartan theory \cite{Trautman:2006} as well as to Poincar\'e gauge theory \cite{Obukhov:2006,Obukhov:2018:1}. Interestingly, Synge was apparently the first who derived the deviation equation for the Riemann-Cartan geometry \cite{Synge:1928}.
 
\subsection{World function and deviation equation}\label{sec_world_dev}

\begin{figure}
  \begin{minipage}[c]{0.55\textwidth}
    \includegraphics[width=\textwidth,angle=-90]{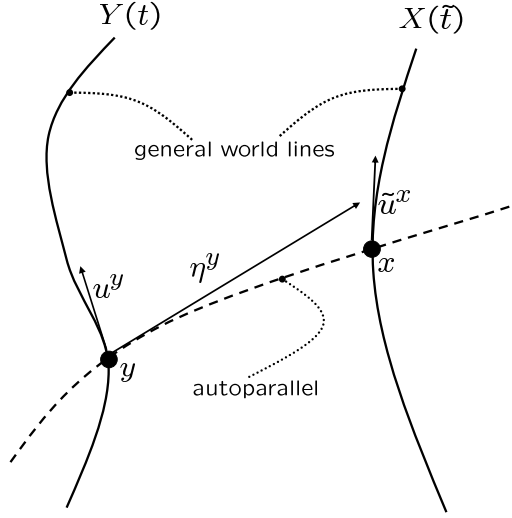}
  \end{minipage}\hfill
  \begin{minipage}[c]{0.4\textwidth}
 \caption{\label{fig_setup} Sketch of the two arbitrarily parametrized world lines $Y(t)$ and $X(\tilde{t})$, and the (dashed) autoparallel connecting two points on these world line. The generalized deviation vector along the reference world line $Y$ is denoted by $\eta^y$.}
  \end{minipage}
\end{figure}

Let us briefly recapitulate the relevant steps which lead to the generalized deviation equation: We want to compare two general curves $Y(t)$ and $X(\tilde{t})$ in an arbitrary spacetime manifold. Here $t$ and $\tilde{t}$ are general parameters, i.e.\ not necessarily the proper time on the given curves. In contrast to the Riemannian case, see section \ref{subsec_theo_found_1}, we now connect two points $x\in X$ and $y\in Y$ on the two curves by the autoparallel joining the two points (we assume that this autoparallel is unique). An autoparallel is a curve along which the velocity vector is transported parallel to itself with respect to the connection on the spacetime manifold. In a Riemannian space autoparallel curves coincide with geodesic lines. Along the autoparallel we have the world function $\sigma$, and conceptually the closest object to the connecting vector between the two points is the covariant derivative of the world function, denoted at the point $y$ by $\sigma^y$, cf.\ figure \ref{fig_setup}. 

%
%
%\begin{figure}
%\begin{center}
%\includegraphics[width=6.5cm,angle=-90]{devrc_fig1.eps}
%\end{center}
%\caption{\label{fig_setup} Sketch of the two arbitrarily parametrized world lines $Y(t)$ and $X(\tilde{t})$, and the (dashed) autoparallel connecting two points on these world line. The generalized deviation vector along the reference world line $Y$ is denoted by $\eta^y$.}
%\end{figure}

At this point, the generality of our derivation of the deviation equation from section \ref{subsec_theo_found_1} pays of, i.e.\ the exact deviation equation given in equation \ref{eta_2nd_deriv_alternative_2} can be directly used in the case in which the connecting curve is an autoparallel therefore, to recapitulate, we have: 
\begin{eqnarray}
\frac{D^2}{dt^2} \eta^{y_1} &=& - \sigma^{y_1}{}_{y_2 y_3} u^{y_2} u^{y_3} - \sigma^{y_1}{}_{y_2} a^{y_2} - \sigma^{y_1}{}_{x_2} \tilde{a}^{x_2} \left(\frac{d\tilde{t}}{dt} \right)^2  \nonumber \\
&&- 2 \sigma^{y_1}{}_{y_2 x_3} u^{y_2} \left( K^{x_3}{}_{y_4} u^{y_4} - H^{x_3}{}_{y_4} \frac{D \sigma^{y_4}}{dt} \right) \nonumber \\
&& - \sigma^{y_1}{}_{x_2 x_3}  \left( K^{x_2}{}_{y_4} u^{y_4} - H^{x_2}{}_{y_4} \frac{D \sigma^{y_4}}{dt} \right) \nonumber \\
&& \times \left( K^{x_3}{}_{y_5} u^{y_5} - H^{x_3}{}_{y_5} \frac{D \sigma^{y_5}}{dt} \right) \nonumber \\
&& - \sigma^{y_1}{}_{x_2} \frac{d t}{d \tilde{t}} \frac{d^2\tilde{t}}{dt^2} \left( K^{x_2}{}_{y_3} u^{y_3} - H^{x_2}{}_{y_3} \frac{D \sigma^{y_3}}{dt} \right). \label{eta_2nd_deriv_alternative_2_rc}
\end{eqnarray}
Again the factor $d \tilde{t} / dt $ by requiring that the velocity along the curve $X$ is normalized, i.e.\ $\tilde{u}^x \tilde{u}_x =1$, in which case we have
\begin{eqnarray}
\frac{d\tilde{t}}{dt} &=& \tilde{u}_{x_1} K^{x_1}{}_{y_2} u^{y_2} -  \tilde{u}_{x_1} H^{x_1}{}_{y_2} \frac{D \sigma^{y_2}}{dt}. \label{formal_vel_3_rc}
\end{eqnarray} 
Equation (\ref{eta_2nd_deriv_alternative_2_rc}) is the exact generalized deviation equation, it is completely general and can be viewed as the extension of the standard geodesic deviation (Jacobi) equation to any order. 

\subsection{World function in Riemann-Cartan spacetime}\label{rc_world_function_sec}

In order to arrive at an expanded approximate version of the deviation equation, we need to work out the properties of a world function based on autoparallels in a Riemann-Cartan background. In contrast to a Riemannian spacetime a Riemann-Cartan spacetime is endowed with an asymmetric connection $\Gamma_{ab}{}^c$, and there will be differences when it comes to the basic properties of a world function $\sigma$ based on autoparallels. We base our presentation on \cite{Puetzfeld:Obukhov:2018:2}, other relevant references which contain some results in a Riemann-Cartan context are \cite{Goldthorpe:1980,NiehYan:1982,Barth:1987,Yajima:1996,Manoff:2000:1,Manoff:2001:1,Iliev:2003:1,Hoogen:2017}. 

For a world function $\sigma$ based on autoparallels, we have the following basic relations in the case of spacetimes with asymmetric connections:
\begin{eqnarray}
\sigma^x \sigma_x = \sigma^y \sigma_y &=& 2 \sigma, \label{rel_1}\\
\sigma^{x_2} \sigma_{x_2}{}^{x_1} &=& \sigma^{x_1}, \label{rel_2}\\
\sigma_{x_1 x_2} - \sigma_{x_2 x_1} &=& T_{x_1 x_2}{}^{x_3} \partial_{x_3} \sigma.  \label{sigma_commutator}
\end{eqnarray}
Note in particular the change in (\ref{sigma_commutator}) due to the presence of the spacetime torsion $T_{x_1 x_2}{}^{x_3}$, which leads to $\sigma_{x_1 x_2} \neq \sigma_{x_2 x_1}$, in contrast to the symmetric Riemannian case, in which $\overline{\sigma}_{x_1 x_2} \stackrel{\rm s}{=} \overline{\sigma}_{x_2 x_1}$ holds\footnote{We use ``s'' to indicate relations which only hold for symmetric connections and denote Riemannian objects by the overbar.}.

In many calculations the limiting behavior of a bitensor $B_{\dots}(x,y)$ as $x$ approaches the references point $y$ is required. This so-called coincidence limit of a bitensor $B_{\dots}(x,y)$ is a tensor 
\begin{eqnarray}
\left[B_{\dots} \right] = \lim\limits_{x\rightarrow y}\,B_{\dots}(x,y),\label{coin}
\end{eqnarray}
at $y$ and will be denoted by square brackets. In particular, for a bitensor $B$ with arbitrary indices at different points (here just denoted by dots), we have the rule \cite{Synge:1960}
\begin{eqnarray}
\left[B_{\dots} \right]_{;y} = \left[B_{\dots ; y} \right] + \left[B_{\dots ; x} \right]. \label{synges_rule}
\end{eqnarray}
We collect the following useful identities for the world function $\sigma$:
\begin{eqnarray}
{}[\sigma]=[\sigma_x]=[\sigma_y]&=&0 , \label{coinc_1}\\
{}[\sigma_{x_1 x_2}]=[\sigma_{y_1 y_2}]&=&g_{y_1 y_2} , \label{coinc_2}\\
{}[\sigma_{x_1 y_2}]=[\sigma_{y_1 x_2}]&=&-g_{y_1 y_2} , \label{coinc_3}\\
{}[\sigma_{x_3 x_1 x_2}] + [\sigma_{x_2 x_1 x_3}] &=&0. \label{coinc_4}
\end{eqnarray}
Note that up to the second covariant derivative the coincidence limits of the world function match those in spacetimes with symmetric connections. However, at the next (third) order the presence of the torsion leads to
\begin{eqnarray}
{}[\sigma_{x_1 x_2 x_3}] &=& \frac{1}{2} \left(T_{y_1 y_3 y_2} + T_{y_2 y_3 y_1}+ T_{y_1 y_2 y_3} \right) = K_{y_2 y_1 y_3} ,
\end{eqnarray}
where in the last line we made use of the contortion\footnote{The contortion $K_{y_2 y_1 y_3}$ should not be confused with the Jacobi propagator $K^x{}_y$.} $K_{ab}{}^c = \overline{\Gamma}{}_{ab}{}^c - \Gamma_{ab}{}^c$. With the help of (\ref{synges_rule}) we obtain for the other combinations with three indices:
\begin{eqnarray}
{}[\sigma_{x_1 x_2 y_3}] &=& -[\sigma_{x_1 x_2 x_3}] =  [\sigma_{y_1 x_2 y_3}] = [\sigma_{x_2 x_3 x_1}] = K_{y_2 y_3 y_1},  \nonumber\\
{}[\sigma_{x_1 y_2 y_3}] &=& -[\sigma_{x_2 x_3 x_1}] = [\sigma_{x_1 x_3 x_2}] = [\sigma_{y_1 x_2 x_3}] = K_{y_3 y_1 y_2},  \nonumber\\
{}[\sigma_{y_1 y_2 y_3}] &=& -[\sigma_{x_3 x_2 x_1}] = K_{y_2 y_1 y_3},  \nonumber\\
{}[\sigma_{x_1 y_2 x_3}] &=& -[\sigma_{x_1 x_3 x_2}] = K_{y_3 y_2 y_1},  \nonumber\\
{}[\sigma_{y_1 y_2 x_3}] &=& \phantom{-}[\sigma_{x_3 x_2 x_1}] = K_{y_2 y_3 y_1}.  \label{coinc_5}
\end{eqnarray}
The non-vanishing of these limits leads to added complexity in subsequent calculations compared to the Riemannian case. 

At the fourth order we have
\begin{eqnarray}
&& K_{y_1}{}^{y}{}_{y_2} K_{y_3 y y_4} + K_{y_1}{}^{y}{}_{y_3} K_{y_2 y y_4} + K_{y_1}{}^{y}{}_{y_4} K_{y_2 y y_4} \nonumber \\
&&+[\sigma_{x_4 x_1 x_2 x_3}] + [\sigma_{x_3 x_1 x_2 x_4}] + [\sigma_{x_2 x_1 x_3 x_4}] =0, \label{coinc_6}\end{eqnarray}
and in particular
\begin{eqnarray}
{}[\sigma_{x_1 x_2 x_3 x_4}] &=&\frac{1}{3}  \nabla_{y_1} \left( K_{y_3 y_2 y_4} + K_{y_4 y_2 y_3}\right) + \frac{1}{3} \nabla_{y_3} \left( 3 K_{y_2 y_1 y_4}  - K_{y_1 y_2 y_4}\right)  
\nonumber \\
&&+ \frac{1}{3} \nabla_{y_4} \left( 3 K_{y_2 y_1 y_3} -  K_{y_1 y_2 y_3}\right) + \pi_{y_1 y_2 y_3 y_4},\label{coinc_11} \\
%---
{}[\sigma_{x_1 x_2 x_3 y_4}] &=& - \frac{1}{3}  \nabla_{y_1} \left( K_{y_3 y_2 y_4} + K_{y_4 y_2 y_3}\right) - \frac{1}{3} \nabla_{y_3} \left( 3 K_{y_2 y_1 y_4}  - K_{y_1 y_2 y_4}\right) \nonumber \\
&&+ \frac{1}{3} \nabla_{y_4} K_{y_1 y_2 y_3} - \pi_{y_1 y_2 y_3 y_4}, \label{coinc_12} \\
%---
{}[\sigma_{x_1 x_2 y_3 y_4}] &=& \frac{1}{3}  \nabla_{y_1} \left( K_{y_4 y_2 y_3} + K_{y_3 y_2 y_4}\right) - \frac{1}{3} \nabla_{y_4} K_{y_1 y_2 y_3}  \nonumber \\
&& - \frac{1}{3} \nabla_{y_3} K_{y_1 y_2 y_4} + \pi_{y_1 y_2 y_4 y_3},  \label{coinc_13}
\end{eqnarray}
\begin{eqnarray}
%--
{}[\sigma_{x_1 y_2 y_3 y_4}] &=&   - \frac{1}{3} \nabla_{y_1} \left( K_{y_3 y_4 y_2} + K_{y_2 y_4 y_3}\right)   + \frac{1}{3} \nabla_{y_3} K_{y_1 y_4 y_2}  + \frac{1}{3} \nabla_{y_2} K_{y_1 y_4 y_3} \nonumber \\
&&+  \nabla_{y_4} K_{y_3 y_1 y_2} - \pi_{y_1 y_4 y_3 y_2},\label{coinc_14}\\ 
%---
{}[\sigma_{y_1 y_2 y_3 y_4}] &=&  \frac{1}{3} \nabla_{y_4} \left( -2 K_{y_2 y_3 y_1} + K_{y_1 y_3 y_2}\right)   - \frac{1}{3} \nabla_{y_2} K_{y_4 y_3 y_1} - \frac{1}{3} \nabla_{y_1} K_{y_4 y_3 y_2} \nonumber \\
&&- \nabla_{y_3} K_{y_2 y_4 y_1} + \pi_{y_4 y_3 y_2 y_1}, \label{coinc_15} \\
\pi_{y_1 y_2 y_3 y_4} &:=& \frac{1}{3} \Big[ K_{y_1 y_2}{}^{y} \left( K_{y_3 y_4 y} + K_{y_4 y_3 y} \right) -  K_{y_1 y_3}{}^{y} \left( K_{y_4 y_2 y} + K_{y y_2 y_4} \right) \nonumber \\
&& - K_{y_1 y_4}{}^{y} \left( K_{y_3 y_2 y} + K_{y y_2 y_3} \right) - 3 K_{y_2 y_1}{}^{y} K_{y_3 y_4 y} + K_{y_3 y_1}{}^{y} K_{y y_2 y_4}  \nonumber \\
&& + K_{y_4 y_1}{}^{y} K_{y y_2 y_3} + R_{y_1 y_3 y_2 y_4} + R_{y_1 y_4 y_2 y_3} \Big]. \label{pi_def}
\end{eqnarray}
Again, we note the added complexity compared to the Riemannian case, in which we have $[\sigma_{x_1 x_2 x_3 x_4}] \stackrel{\rm s}{=} \frac{1}{3} \left(\overline{R}_{y_2 y_4 y_1 y_3} +  \overline{R}_{y_3 y_2 y_1 y_4} \right)$ at the fourth order. In particular, we observe the occurrence of derivatives of the contortion in (\ref{coinc_11})-(\ref{coinc_15}).

Finally, let us collect the basic properties of the so-called parallel propagator $g^{y}{}_{x}:=e^{y}_{(a)} e^{(a)}_{x}$, defined in terms of a parallelly propagated tetrad $e^{y}_{(a)}$, which in turn allows for the transport of objects, i.e.\ $V^y=g^y{}_x V^x, \quad  V^{y_1y_2}=g^{y_1}{}_{x_1} g^{y_2}{}_{x_2} V^{x_1x_2}$, etc., along an autoparallel: 
\begin{eqnarray}
g^{y_1}{}_{x} g^{x}{}_{y_2}&=&\delta^{y_1}{}_{y_2}, \quad g^{x_1}{}_{y} g^{y}{}_{x_2}=\delta^{x_1}{}_{x_2}, \label{parallel_1} \\
%---
\sigma^x \nabla_x g^{x_1}{}_{y_1} &=& \sigma^y \nabla_y g^{x_1}{}_{y_1}=0, \nonumber \\
\sigma^x \nabla_x g^{y_1}{}_{x_1} &=& \sigma^y \nabla_y g^{y_1}{}_{x_1}=0, \label{parallel_2} \\
%---
\sigma_x&=&-g^y{}_x \sigma_y, \quad \sigma_y=-g^x{}_y \sigma_x. \label{parallel_3} 
\end{eqnarray}
Note in particular the coincidence limits of its derivatives
\begin{eqnarray}
\left[g^{x_0}{}_{y_1} \right] &=& \delta^{y_0}{}_{y_1},  \\
\left[g^{x_0}{}_{y_1 ; x_2} \right] &=& \left[g^{x_0}{}_{y_1 ; y_2} \right] = 0, \label{parallel_4} \\
%---
\left[g^{x_0}{}_{y_1 ; x_2 x_3} \right] &=& - \left[g^{x_0}{}_{y_1 ; x_2 y_3} \right] = \left[g^{x_0}{}_{y_1 ; x_2 x_3} \right] \nonumber \\
&=& - \left[g^{x_0}{}_{y_1 ; y_2 y_3} \right] = \frac{1}{2} R^{y_0}{}_{y_1 y_2 y_3}. \label{parallel_5}
\end{eqnarray}

In the next section we will derive an expanded approximate version of the deviation equation. For this we first work out the expanded version of quantities around the reference world line $Y$. In particular, we make use of the covariant expansion technique \cite{Synge:1960,Poisson:etal:2011} on the basis of the autoparallel world function.

\paragraph{Expanded Riemann-Cartan deviation equation}\label{sec_rc_dev}

For a general bitensor $B_{\dots}$ with a given index structure, we have the following general expansion, up to the third order (in powers of $\sigma^y$):
\begin{eqnarray}
B_{y_1 \dots y_n}&=&A_{y_1 \dots y_n} +  A_{y_1 \dots y_{n+1}} \sigma^{y_{n+1}} \nonumber \\
&& + \frac{1}{2} A_{y_1 \dots y_{n+1} y_{n+2}} \sigma^{y_{n+1}} \sigma^{y_{n+2}} + {\cal O}\left( \sigma^3 \right), \label{expansion_general_yn_1} \\
%---
A_{y_1 \dots y_n}&:=&\left[B_{y_1 \dots y_n}\right] , \label{expansion_general_yn_2} \\
%---
A_{y_1 \dots y_{n+1}}&:=&\left[B_{y_1 \dots y_n ; y_{n+1}}\right] - A_{y_1 \dots y_n ; y_{n+1}} , \label{expansion_general_yn_3} \\
%---
A_{y_1 \dots y_{n+2}}&:=&\left[B_{y_1 \dots y_n ; y_{n+1} y_{n+2}}\right]- A_{y_1 \dots y_n y_0} \left[\sigma^{y_0}{}_{y_{n+1} y_{n+2}}\right] \nonumber \\ 
&&  - A_{y_1 \dots y_n ; y_{n+1} y_{n+2}} - 2 A_{y_1 \dots y_n (y_{n+1} ; y_{n+2})} . \label{expansion_general_yn_4}
\end{eqnarray}
With the help of (\ref{expansion_general_yn_1}) we are able to iteratively expand any bitensor to any order, provided the coincidence limits entering the expansion coefficients can be calculated. The expansion for bitensors with mixed index structure can be obtained from transporting the indices in (\ref{expansion_general_yn_1}) by means of the parallel propagator. 

In order to develop an approximate form of the generalized deviation equation (\ref{eta_2nd_deriv_alternative_2_rc}) up to the second order, we need the following expansions of the derivatives of the world function:
\begin{eqnarray}
\sigma_{y_1 y_2}&=&g_{y_1 y_2} + K_{y_2 y_1 y_3} \sigma^{y_3} + {\cal O}\left( \sigma^2 \right),\label{expansion_explicit_sigma_yy} \\
%---
\sigma_{y_1 x_2}&=&-g_{y_1 x_2} + g_{x_2}{}^y K_{y_3 y y_1} \sigma^{y_3} + {\cal O}\left( \sigma^2 \right),\label{expansion_explicit_sigma_xy} \\
%---
\sigma_{y_1 y_2 y_3}&=&  K_{y_2 y_1 y_3} + \frac{1}{3} \Bigg[ \nabla_{y_4} \left( K_{y_2 y_3 y_1} + K_{y_1 y_3 y_2}\right)  \nonumber \\
&-&  \nabla_{y_2} K_{y_4 y_3 y_1} -  \nabla_{y_1} K_{y_4 y_3 y_2} - 3 \nabla_{y_3} K_{y_2 y_4 y_1} \nonumber \\
&+& 3 \pi_{y_4 y_3 y_2 y_1} \Bigg] \sigma^{y_4} + {\cal O}\left( \sigma^2 \right),  \label{expansion_explicit_sigma_yyy}\\
%---
\sigma_{y_1 y_2 x_3}&=& g_{x_3}{}^{y_3} \Bigg\{ K_{y_2 y_3 y_1} - \frac{1}{3} \Bigg[ \nabla_{y_3} \left(K_{y_2 y_4 y_1} + K_{y_1 y_4 y_2}\right)  \nonumber \\
&-& \nabla_{y_2} K_{y_3 y_4 y_1} - \nabla_{y_1} K_{y_3 y_4 y_2} \nonumber \\
&+& 3 \pi_{y_3 y_4 y_2 y_1} \Bigg] \sigma^{y_4}\Bigg\} + {\cal O}\left( \sigma^2 \right) , \label{expansion_explicit_sigma_yyx} 
\end{eqnarray}
\begin{eqnarray}
%---
\sigma_{y_1 x_2 x_3}&=& g_{x_2}{}^{y_2} g_{x_3}{}^{y_3} \Bigg\{ K_{y_3 y_1 y_2} \nonumber \\
&+& \frac{1}{3} \Bigg[ \nabla_{y_2} \left(K_{y_4 y_3 y_1} + K_{y_1 y_3 y_4}\right)  \nonumber \\
&-& \nabla_{y_4} \left( K_{y_2 y_3 y_1} + 3 K_{y_3 y_1 y_2} \right) \nonumber \\
&-& \nabla_{y_1} K_{y_2 y_3 y_4} + 3 \pi_{y_2 y_3 y_4 y_1} \Bigg] \sigma^{y_4} \Bigg\} + {\cal O}\left( \sigma^2 \right). \label{expansion_explicit_sigma_yxx}
\end{eqnarray}
The Jacobi propagators are approximated as follows
\begin{eqnarray}
%---
H^{x_1}{}_{y_2} & =& g^{x_1}{}_{y_2} + K_{y_3 y_2}{}^{x_1} \sigma^{y_3}+ {\cal O}\left( \sigma^2 \right) , \label{jacobi_prop_1_explicit_xy} \\
%---
 K^{x_1}{}_{y_2}& =& g^{x_1}{}_{y_2} + \left(K_{y_2}{}^{x_1}{}_{y_3}+K_{y_3 y_2}{}^{x_1} \right) \sigma^{y_3} + {\cal O}\left( \sigma^2 \right) ,  \label{jacobi_prop_2_explicit_xy}
%---
\end{eqnarray}
which in turn allows for an expansion of the recurring term entering (\ref{eta_2nd_deriv_alternative_2_rc}):
\begin{eqnarray}
K^{x_1}{}_{y_2} u^{y_2} \!-\! H^{x_1}{}_{y_2} \frac{D \sigma^{y_2}}{dt} = g^{x_1}{}_{y'} \Bigl(\! u^{y'} -  \frac{D \sigma^{y'}}{dt} 
+ T_{y_2 y_3}{}^{y'} u^{y_2} \sigma^{y_3} \! \Bigr)  \!+\! {\cal O}\left( \sigma^2 \right)\!. \label{recurrent_term_explicit}
\end{eqnarray} 

\paragraph{Synchronous parametrization}

Before writing down the expanded version of the generalized deviation equation, we will simplify the latter by choosing a proper parametrization of the neighboring curves. The factors with the derivatives of the parameters $t$ and $\tilde{t}$ appear in (\ref{eta_2nd_deriv_alternative_2_rc}) due to the non-synchronous parametrization of the two curves. It is possible to make things simpler by introducing the synchronization of parametrization. Namely, we start by rewriting the velocity as
\begin{equation}
u^y = {\frac {dY^y}{dt}} = {\frac {d\tilde{t}}{dt}} {\frac {dY^y}{d\tilde{t}}}.\label{ut1}
\end{equation}
That is, we now parametrize the position on the first curve by the same variable $\tilde{t}$ that is used on the second curve. Accordingly, we denote 
\begin{equation}
\ut^y = {\frac {dY^y}{d\tilde{t}}}.\label{ut2}
\end{equation} 
By differentiation, we then derive
\begin{eqnarray}
a^y &=& {\frac {d^2\tilde{t}}{dt^2}}\ut^y + \left({\frac {d\tilde{t}}{dt}}\right)^2\at^y,\label{at1}
\end{eqnarray}
where 
\begin{equation}
\at^y = {\frac {D}{d\tilde{t}}}\ut^y = {\frac {D^2Y^y}{d\tilde{t}^2}}.\label{at2}
\end{equation}
Analogously, we derive for the derivative of the deviation vector
\begin{equation}
{\frac {D^2\eta^y}{dt^2}} =  {\frac {d^2\tilde{t}}{dt^2}}{\frac {D\eta^y}{d\tilde{t}}}
+ \left({\frac {d\tilde{t}}{dt}}\right)^2 {\frac {D^2\eta^y}{d\tilde{t}^2}}.\label{Deta2}
\end{equation}
Now everything is synchronous in the sense that both curves are parametrized by $\tilde{t}$.

As a result, the exact deviation equation (\ref{eta_2nd_deriv_alternative_2_rc}) is recast into a simpler form
\begin{eqnarray}
\frac{D^2}{d\tilde{t}^2} \eta^{y_1} &=& - \sigma^{y_1}{}_{y_2} \at^{y_2} - \sigma^{y_1}{}_{x_2} \tilde{a}^{x_2} - \sigma^{y_1}{}_{y_2 y_3} \ut^{y_2} \ut^{y_3} \nonumber \\
&&- 2 \sigma^{y_1}{}_{y_2 x_3} \ut^{y_2} \left( K^{x_3}{}_{y_4} \ut^{y_4} - H^{x_3}{}_{y_4} \frac{D \sigma^{y_4}}{d\tilde{t}} \right) \nonumber \\
&& - \sigma^{y_1}{}_{x_2 x_3}  \left( K^{x_2}{}_{y_4} \ut^{y_4} - H^{x_2}{}_{y_4} \frac{D \sigma^{y_4}}{d\tilde{t}} \right) \nonumber \\
&& \times \left( K^{x_3}{}_{y_5} \ut^{y_5} - H^{x_3}{}_{y_5} \frac{D \sigma^{y_5}}{d\tilde{t}} \right). \label{deveq_master}
\end{eqnarray}

\paragraph{Explicit expansion of the deviation equation}

Substituting the expansions (\ref{expansion_explicit_sigma_yy})-(\ref{recurrent_term_explicit}) into (\ref{deveq_master}), we obtain the final result
\begin{eqnarray}
\frac{D^2}{d\tilde{t}^2} \eta^{y_1} &=& \tilde{a}^{y_1} - \at^{y_1} + T_{y_2y_3}{}^{y_1}\ut^{y_2}\frac{D \eta^{y_3}}{d\tilde{t}} 
- \Bigl(\Delta^{y_1}{}_{y_2y_3y_4}\ut^{y_2}\ut^{y_3} \nonumber \\
&&+\,K_{y_2y_4}{}^{y_1}\at^{y_2} - K_{y_4y_2}{}^{y_1}\tilde{a}^{y_2}\Bigr)\,\eta^{y_4} + {\cal O}\left( \sigma^2 \right),\label{final_explicit}
\end{eqnarray}
where we introduced the abbreviation
\begin{eqnarray}
\Delta_{y_1y_2y_3y_4} &:=& 2 \pi_{y_3y_4y_2y_1} - \pi_{y_4y_3y_2y_1} - \pi_{y_2y_3y_4y_1} + T_{y'y_2y_1}T_{y_4y_3}{}^{y'} \nonumber \\
&&- 2 \nabla_{y_2}K_{(y_1y_3)y_4} + \nabla_{y_1}K_{y_2y_3y_4} - \nabla_{y_4}K_{y_2y_3y_1}.\label{Delta_explicit}
\end{eqnarray}
It should be understood that the last expression is contracted with $\ut^{y_2}\ut^{y_3}$ and hence the symmetrization is naturally imposed on the indices $(y_2y_3)$. 

Equation (\ref{final_explicit}) allows for the comparison of two general world lines in Riemann-Cartan spacetime, which are not necessarily geodetic or autoparallel. It therefore represents the generalization of the deviation equation derived in equation (\ref{eta_2nd_deriv_compact_2}).
 
\paragraph{Riemannian case}

A great simplification is achieved in a Riemannian background, when
\begin{eqnarray}
  \overline{\Delta}_{y_1y_2y_3y_4} &=& 2 \overline{\pi}_{y_3y_4y_2y_1} - \overline{\pi}_{y_4y_3y_2y_1} - \overline{\pi}_{y_2y_3y_4y_1} \nonumber \\
  &=& \overline{R}_{y_1y_3y_2y_4}, \label{Delta_riem} 
\end{eqnarray}
and (\ref{final_explicit}) is reduced to
\begin{eqnarray}
\frac{D^2}{d\tilde{t}^2} \eta^{y_1} &\stackrel{\rm s}{=}& \tilde{a}^{y_1} - \at^{y_1} - \overline{R}^{y_1}{}_{y_2y_3y_4} \ut^{y_2}\ut^{y_3} \eta^{y_4} + {\cal O}\left( \sigma^2 \right). \label{final_riem}
\end{eqnarray}
Along geodesic curves, this equation is further reduced to the well-known geodesic deviation (Jacobi) equation. 

\paragraph{Choice of coordinates}\label{sec_coordinate_choice}

In order to utilize the deviation equation for measurements or in a gravitational compass setup \cite{Synge:1960,Szekeres:1965,Ciufolini:Demianski:1986,Puetzfeld:Obukhov:2016:1}, the occurring covariant total derivatives need to be rewritten and an appropriate coordinate choice needs to be made. The left-hand side of the deviation equation takes the form:
\begin{eqnarray}
\frac{D^2 \eta_a}{dt^2} &=& \dot{u}^b \nabla_b \eta_a +\stackrel{\circ \circ}{\eta}_a - 2 u^b \Gamma_{ba}{}^d \stackrel{\circ}{\eta}_d - u^b u^c \Gamma_{cb}{}^d \partial_d \eta_{a} \nonumber\\
&& - u^b u^c \eta_e \left(\partial_c\Gamma_{ba}{}^e - \Gamma_{cb}{}^d \Gamma_{da}{}^e - \Gamma_{ca}{}^d \Gamma_{bd}{}^e \right).  \label{2nd_deriv_expanded}
\end{eqnarray}
Here we used $\stackrel{\circ}{\eta}\!{}^a:=d\eta^a/dt$ for the standard total derivative. 

Observe that the first term on the right-hand side vanishes in the case of autoparallel curves ($\dot{u}^a:=Du^a/dt=0$). Also note the symmetrization of the connection imposed by the velocities in some terms. 

Rewriting the connection in terms of the contortion and switching to normal coordinates \cite{Veblen:1922,Veblen:Thomas:1923,Thomas:1934,Schouten:1954,Avramidi:1991,Avramidi:1995,Petrov:1969} along the world line, which we assume to be an autoparallel, yields
\begin{eqnarray}
&&\frac{D^2 \eta_a}{dt^2} \stackrel{|_Y}{=} \stackrel{\circ \circ}{\eta}_a + 2 u^b K_{ba}{}^d \stackrel{\circ}{\eta}_d + u^b u^c K_{cb}{}^d \partial_d \eta_{a} \nonumber\\
&& + u^b u^c \eta_e \left(\partial_c K_{ba}{}^e - \frac{2}{3} \overline{R}_{c(ba)}{}^e + K_{cb}{}^d K_{da}{}^e - K_{ca}{}^d K_{bd}{}^e \right).  \label{2nd_deriv_expanded_autoparallel_normal}
\end{eqnarray}
Note the appearance of a term containing the partial (not ordinary total) derivative of the deviation vector, in contrast to the Riemannian case. 

The first term in the second line may be rewritten as an ordinary total derivative, i.e.\ $u^b u^c \eta_e \partial_c K_{ba}{}^e = u^b \eta_e {\stackrel{\circ}{K}_{ba}{\!\!}^e}$, but this is still inconvenient when recalling the compass equation, which will contain terms with covariant derivatives of the contortion. 

\subsection{Operational interpretation}

Some thoughts about the operational interpretation of the coordinate choice are in order. In particular, it should be stressed that we did not specify any {\it physical} theory in which the deviation equation (\ref{deveq_master}) should be applied. Or, stated the other way round, the derived deviation equation is of completely {\it geometrical} nature, i.e.\ it describes the change of the deviation vector between points on two general curves in Riemann-Cartan spacetime. From the mathematical perspective, the choice of coordinates should be solely guided by the simplicity of the resulting equation. In this sense, our previous choice of normal coordinates appears to be appropriate. But what about the physical interpretation, or better, the operational realization of such coordinates? 

Let us recall the coordinate choice in General Relativity in a Riemannian background. In this case normal coordinates also have a clear operational meaning, which is related to the motion of structureless test bodies in General Relativity. As is well known, such test bodies move along the geodesic equation. In other words, we could -- at least in principle -- identify a normal coordinate system by the local observation of test bodies. If other external forces are absent, normal coordinates will locally -- where ``locally'' refers to the observers laboratory on the reference world line -- lead to straight line motion of test bodies. In this sense, there is a clear operational procedure for the realization of normal coordinates.

Here we are in a more general situation, since we have not yet specified which gravitational theory we are considering in the geometrical Riemann-Cartan background. The physical choice of a gravity theory will be crucial for the operational realization of the coordinates. Recall the form of the equations of motion for a very large class \cite{Puetzfeld:Obukhov:2014:2,Obukhov:Puetzfeld:2015:1} of gravitational theories, which also allow for additional internal degrees of freedom, in particular for spin. In this case the equations of motion are no longer given by the geodesic equation or, as it is sometimes erroneously postulated in the literature, by the autoparallel equation. In such theories, test bodies exhibit an additional spin-curvature coupling, which leads to non-geodesic motion, even locally. 

In the context of gravitational theories beyond GR, one should therefore be aware of the fact, that for the experimental realization of the normal coordinates, one now has to make sure to use the correct equation of motion and, consequently, the correct type of test body. Taking the example of a theory with spin-curvature coupling, like Einstein-Cartan theory, this would eventually lead to the usage of test bodies with vanishing spin -- since those still move on standard geodesics, and therefore lead to an identical procedure as in the general relativistic case, i.e.\ one adopts coordinates in which the motion of those test bodies becomes rectilinear.  

\subsection{Summary}\label{sec_summary_3}

This concludes or outlook and the generalization of the deviation equation to a Riemann-Cartan geometry. The generalization should serve as a foundation for the test of gravitational theories which make use of post-Riemannian geometrical structures. As we have discussed in detail, the operational usability of the Riemann-Cartan deviation equation differs from the one in a general relativistic context, which was also noticed quite early in \cite{Hehl:1976}. In contrast to the Riemannian case, an algebraic realization of a gravitational compass \cite{Synge:1960,Szekeres:1965,Puetzfeld:Obukhov:2016:1,Ciufolini:Demianski:1986} on the basis of the deviation equation is out of the question due to the appearance of derivatives of the torsion even at the lowest orders. It remains to be shown which additional concepts and assumptions are needed in order to fully realize a gravitational compass in a Riemann-Cartan background.   

\paragraph{Acknowledgements}
This work was supported by the Deutsche Forschungsgemeinschaft (DFG) through the grant PU 461/1-1 (D.P.). The work of Y.N.O. was partially supported by PIER (``Partnership for Innovation, Education and Research'' between DESY and Universit\"at Hamburg) and by the Russian Foundation for Basic Research (Grant No. 16-02-00844-A).

\newpage

\appendix

\section{Directory of symbols}\label{sec_notation}

\begin{table}[!ht]
\begin{center}
\caption{\label{tab_symbols}Directory of symbols.}
\begin{tabular}{ll}
\hline
\hline
Symbol & Explanation\\
\hline
&\\
\hline
\multicolumn{2}{l}{{Geometrical quantities}}\\
\hline
$g_{a b}$ & Metric\\
$\sqrt{-g}$ & Determinant of the metric \\
$\delta^a_b$ & Kronecker symbol \\
$\varepsilon_{abcd}, \varepsilon_{\alpha\beta\gamma}$  & (4D, 3D) Levi-Civita symbol\\
$s$, $\tau$ & Proper time \\
$x^{a}$, $y^{a}$ & Coordinates \\
$\lambda_b{}^{(\alpha)}$ & (Fermi propagated) tetrad \\
$Y(s)$, $X(\tau)$ & (Reference) world line\\
$\xi^a$ & Constants in spatial Fermi coordinates\\
$\overline{\Gamma}_{a b}{}^c$, $\Gamma_{a b}{}^c$ & (Levi-Civita) connection \\
$T_{ab}{}^c$, $K_{ab}{}^c$ & Torsion, contortion\\
${}^{\ast}\Gamma_{a b \dots}{}^c$ & Derivative of connection (normal coordinates)\\
$R_{a b c}{}^d$, $C_{a b c}{}^d$ & Riemann, Weyl curvature \\
$\sigma$ & World function\\
$\eta^y$ & Deviation vector\\
$g^{y_0}{}_{x_0}$ & Parallel propagator\\
$K^{x}{}_{y}, H^{x}{}_{y}$ & Jacobi propagators \\
&\\
\hline
\multicolumn{2}{l}{{Misc}}\\
\hline
$u^a$, $a^b$ & 4-velocity, 4-acceleration\\
$v^\alpha$, $\omega^\alpha$, $V^\alpha$   & (Linear, rotational, combined) 3-velocity\\
$b^\alpha$, $\eta^\alpha$ & Derivative of (linear, rotational) acceleration \\ 
&\\
\hline
\multicolumn{2}{l}{{Operators}}\\
\hline
$\partial_i$, ``$,$'' & Partial derivative \\ 
$\nabla_i$, ``$;$''  & Covariant derivative \\
$\frac{D}{ds} = $``$\dot{\phantom{a}}$'' & Total covariant derivative \\
$\frac{d}{ds} = $``$\stackrel{\circ}{\phantom{a}}$'' & Total  derivative \\
``$[ \dots ]$''& Coincidence limit\\
``$\overline{\phantom{A}}$''& Riemannian object\\
\hline
\hline
\end{tabular}
\end{center}
\end{table}

\begin{table}
\begin{center}
\caption{\label{tab_symbols_cont}Directory of symbols (continued).}
\begin{tabular}{ll}
\hline
\hline
Symbol & Explanation\\
\hline
&\\
\hline
\multicolumn{2}{l}{{Auxiliary quantities (Method 1)}}\\
\hline
${}^{(m,n)}A_a$, ${}^{(m,n,p)}A_a$ & Accelerations of compass constituents \\
$\alpha^{y_0}{}_{y_1 \dots y_n}$, $\beta^{y_0}{}_{y_1 \dots y_n}$, $\gamma^{y_0}{}_{y_1 \dots y_n}$& Expansion coefficients\\
$c_{(m)a}$, $d_{(m)a}$& Constants\\
$\phi^{y_1}{}_{y_2 \dots}$, $\lambda^{y_1}{}_{y_2 \dots}$, $\mu^{y_1}{}_{y_2 \dots}$, & Abbreviations \\
$\Delta_{i_1\dots}$, $\Xi_{i_1\dots}$ &\\
&\\
\hline
\multicolumn{2}{l}{{Auxiliary quantities (Method 2)}}\\
\hline
$C$ & Frequency ratio \\
$A$, $B$, $K_{1,2,3}$ & Abbreviations\\
&\\
\hline
\multicolumn{2}{l}{{Auxiliary quantities (Outlook)}}\\
\hline
$A_{y_1 \dots y_n}$ & Expansion coefficient \\
$\pi_{y_1 y_2 y_3 y_4}$ & Abbreviation \\
\hline
\hline
\end{tabular}
\end{center}
\end{table}

\bibliographystyle{unsrt}
\bibliography{obukhov_puetzfeld_relgeo_proceedings_2016}

\end{document}